\documentclass[12pt]{article}

\usepackage{epsfig, graphicx}       
\def\NB#1\NE{\color{red}#1\color{black}}
\def\OB#1\OE{\color{blue}#1\color{black}}
\def\CB#1||#2\CE{\color{blue}#1\color{red} #2\color{black} }
\usepackage{%
pstricks,%
pst-plot,%
epsf%
}
\usepackage{cite}
\usepackage{graphicx}
\usepackage{array,dcolumn}      

\newcommand{\hs}{\hspace*{0.5cm}}

\newcommand{\be}{\begin{equation}}
\newcommand{\ee}{\end{equation}}
\newcommand{\bea}{\begin{eqnarray}}
\newcommand{\eea}{\end{eqnarray}}
\newcommand{\bary}{\begin{array}}
\newcommand{\eary}{\end{array}}
\newcommand{\bit}{\begin{itemize}}
\newcommand{\eit}{\end{itemize}}
\newcommand{\ben}{\begin{enumerate}}
\newcommand{\een}{\end{enumerate}}
\newcommand{\crn}{\nonumber \\}

\newcommand{\al}{\alpha}

\newcommand{\ga}{\gamma}

\newcommand{\fr}{\frac}

\newcommand{\bc}{\begin{center}}
\newcommand{\ec}{\end{center}}

\newcommand{\nn}{\nonumber}

\newcommand{\ep}{\epsilon}

\begin{document}

 \bc {\Large \bf \textsc{ Higgs revised in supersymmetric economical 3-3-1 model with $B/\mu$-type terms}} \vspace{0.1cm}

\vspace*{1cm} \textbf{D. T. Binh, L. T. Hue, D. T. Huong and H. N.
Long}

\vspace*{0.5cm}

 \emph{Institute of Physics,   Vietnam Academy of Science  and
Technology, 10 Dao Tan, Ba Dinh, Hanoi, Vietnam}\ec

\noindent

\begin{abstract}
We re-investigate the scalar potential and the Higgs sector of the
 supersymmetric economical 3-3-1 model (SUSYE331) in the presence
 of the $B/\mu$ type terms which has many important consequences. First, the model contains no
 massless Higgs fields. Second, we  prove that
 soft mass parameters of Higgses must be at the $\mathrm{SU(3)}_L$
 scale. As a result, the masses of Higgses drift toward
 this scale except one light real neutral Higgs with the mass of
 $m_Z|c_{2\gamma}|$ at the tree level. We also show that there are
 some Higgses containing many properties of the
  Higgses in the minimal supersymmetric standard model
  (MSSM), especially in the neutral Higgs  sector. One exact relation in
  the MSSM, $m^2_{H^{\pm}}=m^2_{A}+m^2_W$,
 is still true in the SUSYE331. Based on this result we make some comments on the lepton flavor
 violating (LFV) decays of these Higgses as one of signatures of new physics in SUSYE331 model which
  may be detected by present colliders.
\end{abstract}

PACS number(s):  12.60.Jv ,  12.60.Fr , 14.80.Cp

\section{\label{intro}Introduction}
The discovery of a new particle by LHC experiments is the most
intriguing event  in both theoretical and experimental current
physics. As founded  by both ATLAS  and CMS
\cite{higgsdicovery1,higgsdicovery2} this new particle, with mass
around 125.5 GeV, carries many properties of the  Higgs boson
predicted by the Standard Model (SM). On the other hand, many
works tried to determine whether this Higgs is really the SM Higgs
or some new Higgs in models beyond the SM
\cite{lowmH,NonSMHiggs1,NonSMHiggs2}. Many properties of this new
Higgs are available  in \cite{CERNreport}. Some very helpful
discussions on which models are  excluded or still acceptable by
the existence of the new Higgs found are for example in \cite{guido}.
At this time apart from the SM, the MSSM is the most attractive
model which both experimental and theoretical physics focus on.
For the review of SM Higgs see \cite{Djouadi2}. A review of MSSM
Higgses is in \cite{Djou,Martin}. For
 the MSSM there are five physical Higgses, including  one CP-odd
neutral Higgs and two CP-even neutral Higgses. The mass of the
lighter neutral Higgs is shown to be smaller than of $m_Z
|\cos(2\beta)|$ at tree level. Here $\beta$ is determined by the
relation $t_\beta=v_2/v_1$, the ratio of  the two Higgs vacuum
expectation values (VEVs) in the MSSM. The  mass of this light Higgs
can increase up to 135 GeV after including loop
corrections \cite{degrassi}. Of course,  the value of 125.5 GeV
 still satisfies this
constraint but the mass spectrum of supersymmetric particles has
drifted to the TeV scale \cite{lowmH,NonSMHiggs2}.

There is another class of supersymmetric (SUSY)  models, called
SUSY 3-3-1 models, which is not mentioned above. The SUSY 3-3-1
models are SUSY versions of the 3-3-1 models \cite{331m,331r}
constructed in order to explain some issues as the so-called
family replication, the electric charged quantization \cite{ecq},
the large difference between masses of quarks in different
families \cite{longvan}... The greatest disadvantage of these models
is the complication in the Higgs sector, namely: these  models need
many Higgs multiplets to generate the masses of the fermions. Some models
with the simplest Higgs sector, such as
\cite{haihiggs,Ferreira,higgseconom} need only two
$\mathrm{SU(3)}_L$ Higgs triplets. But some fermions in these
models get zero
 masses at the tree level and they need to get
non-vanishing  masses from loop corrections \cite{haihiggs} or
effective non-renomalizable  operators \cite{Ferreira}. To solve
this problem as well as the problem of dark matter in these models,
some supersymmetric versions of these 3-3-1 models were introduced
\cite{susyec,s331r,susyrm331}. These models, of course, keep
interesting properties of the 3-3-1 as well as SUSY models. But
the needed Higgs multiplets are doubled compared with non-SUSY
version to cancel the gauge anomaly caused by  Higgsinos. The Higgs
sectors are now much more complicated. Anyway,  they were
investigated in detail for supersymmetric economical 3-3-1
(SUSYE331) model \cite{susyec,Dong3},  supersymmetric reduced
minimal 3-3-1 (SUSYRM331) model \cite{susyrm331}. In this work we
will concentrate on the SUSYE331 Higgs sector for two
important reasons:
\begin{itemize}
    \item First, the SUSYE331 has the simplest Higgs sector in SUSY331
    models and it  was widely investigated as regards phenomenology
    such as Higgs sector \cite{susyec,Dong3}, inflation scenarios
    \cite{inflasusye331}, mass spectrum of SUSY particles
    \cite{susymaspectrum,sfSUSYE331}, and lepton flavor violating (LFV) decays \cite{giang,hue1}. One
    problem  of this model is the absence of
    $B/\mu$-type terms in the scalar potential. These terms are
    very important for the vacuum stability of general SUSY
    models. They were first addressed in \cite{hue1} but the
    consequences of their presence were not shown in
    detail.
    \item Second, as mentioned above, the presence of the 125.5 GeV Higgs
    strongly  affects the parameter space of all present models
    including SUSY331 models. It is indeed necessary to consider
    the reality of the SUSYE331 under the impact the appearance of this Higgs.
\end{itemize}
 Comparing the Higgs sectors of the SUSY331 models with that of
 the  MSSM is the straightforward way to estimate the compatibility of
them  with Higgs experiments at this
 time. For the SUSYE331, we try to identify some Higgses as
 "like-MSSM"  Higgses and the others as being really $\mathrm{SU(3)}_L$
 Higgses. While  finding the exact physical solutions for
 Higgses in the presence of the $B/\mu$-type terms is almost impossible, we
 can calculate them in an approximate way with high
 accuracy, based on the presence  of the
$\mathrm{SU(3)}_L$ scale itself. In the SUSY versions such as
SUSYE331 this scale corresponds to two Higgses $\chi$ and $\chi'$
with two VEVs, which is assumed to be much larger than the SM symmetry
breaking scale $w,w'\gg  246$ GeV. Constraints from the heavy
neutral $Z'$ of the model predict that this scale is of the order of the
 TeV scale \cite{coutinho}. Combining with the conditions of
 the minimum of the scalar potential, it can be deduced that both
 $B/\mu$-terms and soft parameters should be of  the same
 order, the electroweak  $\mathcal{O}(m^2_W)$ or the $\mathrm{SU(3)}_L$
 scale. We  show  this conclusion in detail in Sect.
 \ref{sec:massspectrum}. In
 that section, we also construct
 all squared  Higgs mass matrices of the model, find exact
 solutions for physical CP-odd neutral Higgses, and establish two
 equations determining the mass eigenvalues of CP-even neutral and
 charged Higgses.  Approximate solutions of these Higgs masses will be discussed in Sect.
 \ref{constrainH} after we prove that the  $B/\mu$-terms and soft
 parameters favor the $\mathrm{SU(3)}_L$ scale. With this condition,
 the Higgs spectrum of the SUSYE331  is split into two
   parts, in which the
 first part contains Higgses with properties being similar to MSSM
 Higgses. Some other Higgs properties  are also mentioned in this
 section. Furthermore,  in Sect.
 \ref{mssmvs331}, like-MSSM  Higgses are discussed in  more detail  by comparing them with MSSM  Higgses
  in coupling with the standard particles. In Sect. \ref{LFVHiggs},
   we discuss the LFV decay of the
 neutral Higgs, $H^0\rightarrow \mu\tau $, in the SUSYE331 model.
 This kind of decay was investigated in \cite{giang} without the
 appearance of $B/\mu$-type terms. It is noted that detecting LFV
 decay at TeVatron and LHC was discussed in \cite{assa}, and the
 sensitivity of the LHC for these decays has been discussed
 \cite{lfvlhc}. In the revised SUSYE331
 version, only MSSM-like Higgses can have a large
 LFV decay branching ratio for $H^0\rightarrow \mu\tau $. This result
 is easily obtained  based on many previous works on this kind of
 decay for the MSSM  and extended versions of
  the  MSSM \cite{lfvMSSM,Anna1,Dia1}. First of all,  we start our work
 by reviewing the SUSYE331 particle content in Sect. \ref{model}.
\section{\label{model}A review of the model SUSYE331}
 In this section we only  list the particle content of the
 SUSYE331 which we consider in this work.
 The details  were
 thoroughly investigated for example in \cite{susyec,Dong3}.

 The superfield content is defined in a standard way as
follows:
\be \widehat{F}= (\widetilde{F}, F),\hs \widehat{S} = (S,
\widetilde{S}),\hs \widehat{V}= (\lambda,V), \ee where the
components $F$, $S$ and $V$ stand for the fermion, scalar, and
vector fields, while their superpartners are denoted
$\widetilde{F}$, $\widetilde{S}$ and $\lambda$, respectively
\cite{s331r,susyrm331}.

The superfields containing leptons under the 3-3-1 gauge group
transform as
\begin{equation}
\widehat{L}_{a L}=\left(\widehat{\nu}_{a}, \widehat{l}_{a},
\widehat{\nu}^c_{a}\right)^T_{L} \sim (1,3,-1/3),\hs
  \widehat {l}^{c}_{a L} \sim (1,1,1),\label{l2}
\end{equation} where $\widehat{\nu}^c_L=(\widehat{\nu}_R)^c$ and $a=1,2,3$
is a generation index.

The superfields for the left-handed quarks of the first generation
are in triplets, \be \widehat Q_{1L}= \left(\widehat { u}_1,\
                        \widehat {d}_1,\
                        \widehat {u}^\prime
 \right)^T_L \sim (3,3,1/3).\label{quarks3}\ee
  We omit the color index of quarks. The right-handed singlet counterparts of
  these superfields  are  denoted
  \be
\widehat {u}^{c}_{1L},\ \widehat { u}^{ \prime c}_{L} \sim
(3^*,1,-2/3),\hs \widehat {d}^{c}_{1L} \sim (3^*,1,1/3 ).
\label{l5} \ee Conversely, the last two generations contained in
superfields which transform as antitriplets of the
$\mathrm{SU(3)}_L$
\begin{equation}
\begin{array}{ccc}
 \widehat{Q}_{\alpha L} = \left(\widehat{d}_{\alpha}, - \widehat{u}_{\alpha},
 \widehat{d^\prime}_{\alpha}\right)^T_{L} \sim (3,3^*,0), \hs \al=2,3, \label{l3}
\end{array}
\end{equation}
while the right-handed counterparts are in singlets,
\begin{equation}
\widehat{u}^{c}_{\alpha L} \sim \left(3^*,1,-2/3 \right),\hs
\widehat{d}^{c}_{\alpha L},\ \widehat{d}^{\prime c}_{\alpha L}
\sim \left(3^*,1,1/3 \right). \label{l4}
\end{equation}
The prime superscript is used to distinguish exotic quarks  and SM
quarks having the same electric charges. The mentioned fermion
content is originally from  the 3-3-1 model with right-handed
neutrinos \cite{331r,haihiggs},  so it is anomaly-free.

The two superfields $\widehat{\chi}$ and $\widehat {\rho} $
contain the scalar sector of the economical 3-3-1 model (E331)
\cite{higgseconom}: \bea \widehat{\chi}&=& \left
(\widehat{\chi}^0_1, \widehat{\chi}^-, \widehat{\chi}^0_2
\right)^T\sim (1,3,-1/3),\hs  \widehat{\rho}= \left
(\widehat{\rho}^+_1, \widehat{\rho}^0, \widehat{\rho}^+_2\right)^T
\sim  (1,3,2/3). \label{l8} \eea
 To cancel the chiral anomalies of the
Higgsino sector, two extra superfields $\widehat{\chi}^\prime$ and
$\widehat {\rho}^\prime $ are added as follows:
 \be
\widehat{\chi}^\prime= \left (\widehat{\chi}^{\prime 0}_1,
\widehat{\chi}^{\prime +},\widehat{\chi}^{\prime 0}_2
\right)^T\sim ( 1,3^*,1/3), \;\; \widehat{\rho}^\prime = \left
(\widehat{\rho}^{\prime -}_1,
  \widehat{\rho}^{\prime 0},  \widehat{\rho}^{\prime -}_2
\right)^T\sim (1,3^*,-2/3). \label{l10} \ee

 According to
analysis in \cite{susyec}, at the tree level,  $\rho'$ is enough
to generate masses for all charged leptons, while it contributes in
 part to the down-quarks masses.  Also, the $\rho$ generates masses to the
 neutral leptons and contributes in part to the up-quarks
 masses. On the other hand,  both $\chi$ and $\chi'$ only contribute to masses of
  both usual  and exotic quarks.
  It can be supposed  that $\rho$ and $\rho'$ may play similar roles
  as Higgses in the MSSM. It is recalled that the above Higgs sector does not generate masses
  for all quarks of the model. Therefore corrections from loop levels are needed.

As normal 3-3-1 models, the $ \mathrm{SU}(3)_L \otimes
\mathrm{U}(1)_X$ gauge group is broken via two steps:
 \be \mathrm{SU}(3)_L \otimes
\mathrm{U}(1)_X \stackrel{w,w'}{\longrightarrow}\ \mathrm{SU}(2)_L
\otimes \mathrm{U}(1)_Y\stackrel{v,v',u,u'}{\longrightarrow}
\mathrm{U}(1)_{Q},\label{stages}\ee where the VEVs are defined by
\bea
 \sqrt{2} \langle\chi\rangle^T &=& \left(u, 0, w\right), \hs \sqrt{2}
 \langle\chi^\prime\rangle^T = \left(u^\prime,  0,
 w^\prime\right),\\
\sqrt{2}  \langle\rho\rangle^T &=& \left( 0, v, 0 \right), \hs
\sqrt{2} \langle\rho^\prime\rangle^T = \left( 0, v^\prime,  0
\right).\eea
The vector superfields $\widehat{V}^a_c$, $\widehat{V}^a$ and
$\widehat{V}^\prime$ containing the usual gauge bosons are related
with  gauge groups $\mathrm{SU}(3)_C$, $\mathrm{SU}(3)_L$ and
$\mathrm{U}(1)_X $.

The VEVs $w$ and $w^\prime$
are responsible for first stage of  symmetry breaking,
 $\mathrm{SU(3)}_L\times U(1)_X\rightarrow SU(2)_L \times U(1)_Y$
and that provides the mass for new particles, namely  \bea T_i
\left(0,~0,~\frac{w}{\sqrt{2}}\right)^T &\neq& 0, \hs T_i
\left(0,~0,~\frac{w'}{\sqrt{2}}\right)^T \neq 0, \hs  i =
4,5,6,7,8, \crn  X\left(0,~0,~\frac{w}{\sqrt{2}}\right)^T&\neq& 0,
\hs X \left(0,~0,~\frac{w'}{\sqrt{2}}\right)^T \neq 0. \eea
In the gauge boson sector, only the new gauge bosons  $Y^{\pm},X, X^*$
and $ Z^\prime$ gain masses at this stage of symmetry breaking. In
contrast, the three other generators $T_1,~T_2$ and $T_3$
characterizing the $\mathrm{SU(2)}_L$ group are conserved. Also,
the generator of the $U(1)_Y$, defined as
$$ \frac{Y}{2}=-\frac{1}{\sqrt{3}} T_8+ X, $$ is also conserved. We would like to
emphasize that at the first stage of breaking, there is no mixture
between the $Z$ and the $Z^\prime.$ In the second stage the standard
model electroweak symmetry is broken down to $U(1)_Q$ by $u,\
u^\prime$ and $v,\ v^\prime$ and this is responsible for the masses of the
ordinary particles. To keep consistency with the MSSM, we should
suppose \bea u,u ^\prime, v, v^\prime \ll w, w^\prime. \eea
 For more details, the reader is refered to \cite{susyec}.
After the first step of symmetry breaking, we can obtain the
effective Lagrangian for Higgs fields. From the effective Higgs
potential,  we can proceed with the discussion  by  comparison with the MSSM
Higgs sector.


The full Lagrangian  of the model has the form
$\mathcal{L}_{\mathrm{susy}}+\mathcal{L}_{\mathrm{soft}}$, where
the first term is the supersymmetric part and  the last term explicitly breaks
the supersymmetry. More details of this Lagrangian are
discussed in \cite{susyec}. Our work mainly focuses on the Higgs
sector of the model.
\section{\label{sec:massspectrum} Revised scalar potential  for Higgses and  Higgs
sector}
In the soft term  involving the scalar potential, we add  a new
term,
$$\left(b_{\rho} \rho\rho^{\prime}+b_{\chi}\chi\chi^{\prime}+
\mathrm{H.c}\right), $$ to the original supersymmetric Higgs
potential constructed in \cite{susyec}. The revised potential now
is
   \bea
V_{\mathrm{SUSYE331}} &\equiv & V_{\mathrm{scalar}} +
V_{\mathrm{soft}}\crn
 &=& \frac{\mu_{\chi}^2}{4}\left(\chi^\dagger\chi+\chi^{\prime\dagger}\chi^\prime\right)
+\frac{\mu_{\rho}^2}{4}\left(\rho^\dagger\rho+\rho^{\prime\dagger}\rho^\prime\right)
\crn &&+\frac{g^{\prime2}}{12} \left(-\frac{1}{3} \chi^{\dagger}
\chi + \frac{1}{3} \chi^{\prime \dagger} \chi^{\prime}+
\frac{2}{3} \rho^{\dagger}\rho - \frac{2}{3} \rho^{\prime
\dagger}\rho^{\prime} \right)^2\crn & &
+\frac{g^2}{8}\sum_{b=1}^{8}(\chi^\dagger_i\lambda^b_{ij}\chi_j-
\chi^{\prime\dagger}_i\lambda^{*b}_{ij}\chi^\prime_j+
\rho^\dagger_i\lambda^b_{ij}\rho_j-
\rho^{\prime\dagger}_i\lambda^{*b}_{ij}\rho^\prime_j)^2\!\crn
&&+m^2_\rho\rho^\dagger\rho+m^2_\chi\chi^\dagger\chi
+m^2_{\rho^\prime}\rho^{\prime\dagger}\rho^\prime+
m^2_{\chi^\prime}\chi^{\prime\dagger}\chi^\prime
\crn&&-\left(b_{\rho} \rho\rho^{\prime}+
b_{\chi}\chi\chi^{\prime}+
 \mathrm{H.c.}\right). \label{p4} \eea
As discussed in the MSSM, we can redefine the phases of the Higgs fields
in order to get real values of both $ b_{\chi}$ and $b_{\rho}$. In
addition, these parameters must be positive to avoid the minimum
value of the potential corresponding to the zero values of the neutral
Higgses. It implies that electroweak symmetric breaking does not
occur.

Assuming that the VEVs of neutral components $u,\ u',\ v,\ v',\ w$
and $w'$ are real, we expand all Higgs fields around the VEVs as
follows\bea \chi^T&=&\left(
 \begin{array}{ccc}
            \fr{u+S_1+iA_1}{\sqrt{2}}, & \chi^{-}, & \fr{w+S_2+iA_2}{\sqrt{2}} \\
           \end{array}
         \right), \hs  \rho^T = \left(
 \begin{array}{ccc}
  \rho_1^+, & \fr{v+S_5+iA_5}{\sqrt{2}}, & \rho_2^+ \crn
          \end{array}
      \right),\label{2}\\ {\chi^\prime}^T&=&\left(
           \begin{array}{ccc}
\fr{u^\prime+S_3+iA_3}{\sqrt{2}}, & \chi^{\prime
+}, & \fr{w^\prime+S_4+iA_4}{\sqrt{2}} \\
           \end{array}
         \right),\hs {\rho ^\prime}^T =\left(
                         \begin{array}{ccc}
                           \rho_1^{\prime -}, &
\fr{ v^\prime +S_6+iA_6}{\sqrt{2}}, & \rho_2^{\prime-} \\
\end{array}\right).\label{1}
\eea
 The minimum of the $V_{\mathrm{SUSYE331}}$ is equivalent to  the
  canceling of  five linear neutral Higgs terms,
 as listed below:
\bea \mu_\rho^2+4 m_\rho^2 &=& 4 \frac{v'}{v} b_{\rho} -\frac{2
g^{2 \prime}+9g^2}{27}\left[ 2\left(v^2-v^{2 \prime} \right) +w^{2
\prime}-w^2 +u^{ \prime 2}-u^2\right],\crn
  \mu_\chi^2+ 4 m_\chi^2&=& 4 \frac{u'}{u}b_{\chi} -\frac{ g^{\prime 2}}{27}\left[
w^2-w^{\prime 2}+u^2-u^{\prime 2} + 2\left(v^{\prime 2}-v^2
\right) \right] \nonumber
\\ && -\frac{g^2}{3}\left[2\left( u^2-u^{\prime 2}+w^2-w^{\prime
2}\right) +v^{\prime 2}-v^2 \right], \label{linear2}\eea
 \be m_{\rho}^2+m^2_{\rho \prime} +\frac{1}{2} \mu^2_{\rho} = b_{\rho}~
 \frac{v^2+v^{\prime2}}{v v'},\label{linear3}\ee
 \be m_{\chi}^2+m^2_{\chi \prime} + \frac{1}{2} \mu^2_{\chi} = b_{\chi}~\frac{u^2+u^{\prime 2}}{u
 u'},\label{linear4}\ee
 \be (-u' w+u w')\left[b_{\chi}+ \frac{g^2}{4}(u u'+w w')\right]=0.\label{linear5}\ee

From condition (\ref{linear5}), it is easy to see that we have  the equality
$u/u'=w/w'$, the same as shown in \cite{susyec}. The formulas
in (\ref{linear2}) are obtained when this equality is inserted in
 four other independent linear vanishing conditions.
For convention  we will use the notations defined in previous works,
\bea \tan{\beta}&=&t_\beta=\frac{u}{u'},\hs
\tan\gamma=t_\gamma=\frac{v}{v'},\hs t=\frac{g'}{g},\crn
 m_W^2&=&\frac{g^2}{4}\left(v^2+v^{\prime2}\right), \hs
 m_{X}^2=\frac{g^2}{4}\left(u^{\prime2}+w^{\prime2}\right)
 \left(t^2_{\beta}+1\right),
\label{notation1}\eea
 where $m_X$ and $m_W$ are the masses of the non-Hermitian boson $X$
  and $W$ boson, respectively

 Four Eqs. (\ref{linear2})--(\ref{linear4})  now can be
 rewritten in the form
\be \frac{1}{4}\mu_\rho^2+ m_\rho^2 = \frac{b_{\rho}}{t_{\gamma}}
+\frac{2
t^2+9}{27}\left[-m_X^2\cos2\beta+2m_W^2\cos2\gamma\right],
\label{linear6}\ee
 \be
 \frac{1}{4} \mu_\chi^2+ m_\chi^2=\frac{ b_{\chi}}{t_{\beta}} +\frac{t^2+18}{27}
  m_X^2\cos2\beta-\frac{(2 t^2+9)}{27}
m_W^2\cos2\gamma, \label{linear7}\ee
\be s_{2\gamma}\equiv
\sin2\gamma=\frac{2b_{\rho}}{m_{\rho}^2+m^2_{\rho \prime}
+\frac{1}{2} \mu^2_{\rho}}, \hs
s_{2\beta}\equiv\sin2\beta=\frac{2b_{\chi}}{ m_{\chi}^2+m^2_{\chi
\prime} + \frac{1}{2} \mu^2_{\chi}}. \label{linear8}\ee
The two equations in (\ref{linear8}) directly tell  us two separated
constraints  for  $b_{\rho}$ and $b_{\chi}$
\bea  2 b_{\rho} \leq  m_{\rho}^2+m^2_{\rho \prime} +\frac{1}{2}
\mu^2_{\rho} \hs \mathrm{and} \hs 2b_{\chi}\leq
m_{\chi}^2+m^2_{\chi \prime} + \frac{1}{2}
\mu^2_{\chi}.\label{cofbmuchi}\eea

These two conditions are similar to the constraint to the $b$-term in
the $D$-flat directions of the MSSM.  They guarantee that the scalar
potential has a lower bound. So it will have a minimum.

Using results in (\ref{linear8}) to solve the series of two equations
(\ref{linear6}) and (\ref{linear7})  we can determine
$\cos2\gamma$ and $\cos2\beta$ as functions of soft parameters.
But it will be more convenient  to estimate the order of soft parameters by writing
$\cos2\gamma$ and $\cos2\beta$ as follows:
\bea c_{2\gamma}\equiv \cos2\gamma&=& \frac{2
c^2_W\left(\frac{1}{4}\mu^2_{\rho}+m^2_{\rho}-\frac{b_{\rho}}{t_{\gamma}}\right)
+\left(\frac{1}{4}\mu^2_{\chi}+m^2_{\chi}-\frac{b_{\chi}}{t_{\beta}}\right)}{m^2_W},
\crn
 c_{2\beta}\equiv\cos2\beta&=&
\frac{\left(\frac{1}{4}\mu^2_{\rho}+m^2_{\rho}-\frac{b_{\rho}}{t_{\gamma}}\right)+2
\left(\frac{1}{4}\mu^2_{\chi}+m^2_{\chi}-\frac{b_{\chi}}{t_{\beta}}\right)}{m^2_X}\crn&=&
\frac{2m^2_Wc_{2\gamma}}{m^2_X}-
\frac{(3-4s^2_W)\left(\frac{1}{4}\mu^2_{\rho}+m^2_{\rho}-\frac{b_{\rho}}{t_{\gamma}}\right)}{m^2_X}.
\label{condition1}\eea
It is very important to note that the two equations in (\ref{condition1})
have upper bounds:  $|c_{2\gamma}|, |c_{2\beta}|\leq 1$. Combined
with the property $m_W\ll m_X$ of the SUSYE331, the parameters on
the right hand side of (\ref{condition1})  must be on the same
scale of $\mathcal{O}(m^2_W)$ or $\mathcal{O}(m^2_X)$. It means
that  we have only two cases,
\bea
\left|\frac{1}{4}\mu^2_{\rho}+m^2_{\rho}-\frac{b_{\rho}}{t_{\gamma}}\right|
&\sim&
\left|\frac{1}{4}\mu^2_{\chi}+m^2_{\chi}-\frac{b_{\chi}}{t_{\beta}}\right|
\sim \mathcal{O}(m^2_W),\label{case1}\\
\mathrm{or}\hs
  \left|\frac{1}{4}\mu^2_{\rho}+m^2_{\rho}-\frac{b_{\rho}}{t_{\gamma}}
  \right|&\sim& \left|\frac{1}{4}\mu^2_{\chi}+m^2_{\chi}-\frac{b_{\chi}}{t_{\beta}}\right|
\sim \mathcal{O}(m^2_X).\label{case2}\eea
If there is not much hierarchy among the  soft and $\mu_{\rho,\chi}$
parameters, they all should be of the same scale.  In
addition, the  case of (\ref{case2})  appears when the two quantities
$2c^2_W\left(\frac{1}{4}\mu^2_{\rho}+m^2_{\rho}-\frac{b_{\rho}}{t_{\gamma}}\right)$
and
$\left(\frac{1}{4}\mu^2_{\chi}+m^2_{\chi}-\frac{b_{\chi}}{t_{\beta}}\right)
$ have opposite signs, so that they cancel each other to result the
total being of  the $\mathcal{O} (m^2_W)$ scale. The degeneration
among the supersymmetric parameters characterized for a large breaking
scale also happens in the normal $\mathrm{SU(2)}_L\times
\mathrm{U(1)}_L$ supersymmetric model.

 Because the Higgs sector in this model is very complicated, it
 is not easy to find the exact solutions for the mass spectrum
  as well as the mass eigenstates of Higgses. Instead,  we will use some
  appropriate  approximations  to solve the problems.
  In the next section we will use the parameter $\epsilon=m^2_W/m^2_X$,
 which satisfies  $\epsilon\ll1$, as the perturbative variable to do
 approximate calculations.

 We firstly determine  mass eigenvalues of the pseudo-scalar neutral Higgses
because they are calculated exactly. We will use them as
independent parameters in formulas representing the Higgs mass
spectra.

\subsection{\label{psHIggs}Pseudo scalar or CP-odd neutral Higgses}
 The mass Lagrangian of pseudo-scalar Higgses  is split into two
 parts,
 \bea -\mathcal{L}^{\mathrm{mass}}_{A}&=& \frac{1}{2}(A_1,A_2,A_3,A4)\times M^2_{A\chi}
 (A_1,A_2,A_3,A4)^T\crn&+&\frac{1}{2}(A_5,A_6)
 M^2_{A\rho}(A_5,A_6)^T\label{Pnhigss}\eea
  with
 $$ M^2_{A\chi}=\frac{g^2}{4}\left(%
\begin{array}{cccc}
    w^{\prime2}+\frac{4b_{\chi}}{g^2t_{\beta}},& -u' w',&  w w' +\frac{4b_{\chi}t_{\beta}}{g^2},&
      -u w't_{\beta},\\
  -u' w', & u^{\prime2}+ \frac{4b_{\chi}}{g^2t_{\beta}}  & -u w't_{\beta} &
   u u^{\prime} +\frac{4b_{\chi}t_{\beta}}{g^2} \\
  w w't_{\beta}+\frac{4b_{\chi}}{g^2} & -u w't_{\beta} & w^2t_{\beta}+\frac{4b_{\chi}t^2_{\beta}}{g^2}&
  -u w t_{\beta}^2 \\
  -u w't_{\beta} & u u't_{\beta}+\frac{4b_{\chi}}{g^2} &-u w t_{\beta}^2 & u^2t_{\beta}^2+
  \frac{4b_{\chi}t_{\beta}}{g^2} \\
\end{array}%
\right),$$  and
$$ M^2_{A\rho}=\frac{4b_{\rho}}{t_{\gamma}}\times\left(%
\begin{array}{cc}
 1 & t_{\gamma} \\
  t_{\gamma} &t^2_{\gamma} \\
\end{array}%
\right).$$
This leads to the result that there are three massless solutions
and three massive ones, defined as
 \bea m^2_{A_1}\equiv m^2_{\mathrm{H}_{A1}}&=&\frac{2 b_{\rho}}{s_{
 2\gamma}}=\frac{1}{2}\mu^2_{\rho}+m^2_{\rho}+m^2_{\rho'},\crn
  m^2_{A_2}\equiv m^2_{\mathrm{H}_{A2}}&=& \frac{2b_{\chi}}{s_{2\beta}}
  =\frac{1}{2}\mu^2_{\chi}+m^2_{\chi}+m^2_{\chi'},\crn
   m^2_{A_3}\equiv m^2_{\mathrm{H}_{A3}}&=& m^2_{A_2}+m^2_X.
  \label{masscHiggs}\eea
  Because $\rho$ and $\rho'$  play the roles of MSSM  Higgses, $H_{A_1}$ seems to be the same as
  the CP-odd Higgs in
   the MSSM. To compare Higgs mass spectrum
  with the  $\mathrm{SU(3)}_L$ scale in  the following calculations, we will use some
  new  notations, defined by
  \be k_1=\frac{m^2_{A_1}}{m^2_X},\hs k_2=\frac{m^2_{A_2}}{m^2_X}, \hs h_W=\sqrt{\frac{1}{3-4s^2_W}}.
  \label{notation2}\ee
 It is easy to write three massive eigenstates as
  \bea  H_{A_1}&=& A_5c_{\gamma}+A_6s_{\gamma},\crn
  H_{A_2}&=& A_1c_{\beta}s_{\zeta}+A_2c_{\beta}c_{\zeta}+
  A_3s_{\beta}s_{\zeta} +A_4s_{\beta}c_{\zeta},\crn
  H_{A_3}&=&- A_1c_{\beta}c_{\zeta}+A_2c_{\beta}s_{\zeta}
  -A_3s_{\beta}c_{\zeta}+A_4s_{\beta}s_{\zeta},\label{pmase1}\eea
 where $\tan\zeta=u'/w', \cos\zeta=c_\zeta, \sin\zeta=s_\zeta, \cos \beta=c_\beta, \sin \beta=s_\beta,
  \cos \gamma=c_\gamma, \sin \gamma=s_\gamma$. Three massless eigenstates are
\bea  H_{A_4}&=& -A_5s_{\gamma}+A_6c_{\gamma},\crn
  H_{A_5}&=&-A_2s_{\beta}+A_4c_{\beta} ,\crn
  H_{A_6}&=&- A_1s_{\beta}+A_3c_{\beta}.\label{pmase2}\eea
They are Goldstone bosons eaten by neutral gauge bosons $Z, ~Z'$
and $X^0$. There  do not exist any physical massless CP-odd
neutral Higgses in the model.

\subsection{Neutral scalar Higgs}

In the basis of $(S_1,S_2,S_3,S_4,S_5,S_6)$ the squared mass
matrix of real scalar neutral Higgses can be written in the form
of
 \bea M^2_{6S}
&=& \left(
\begin{array}{cccccc}
m_{S11}^2 & m_{S12}^2& m_{S13}^2 & m_{S14}^2 & m_{S15}^2 & m_{S16}^2\\
& m_{S22}^2 & m_{S23}^2 & m_{S24}^2 & m_{S25}^2 & m_{S26}^2 \\
 &  & m_{S33}^2 & m_{S34}^2 & m_{S35}^2 & m_{S36}^2 \\
 &  &  & m_{S44}^2 & m_{S45}^2 & m_{S46}^2 \\
 &  & &  & m_{S55}^2 & m_{S56}^2 \\
  & & &  &  & m_{S66}^2 \\
 \end{array}\right), \label{higgstrunghoa}\eea
where  precise formulas of elements are listed in the Appendix
\ref{ehiggsmaas}. The squared mass matrices of both neutral and
charged Higgses are different from those in \cite{susyec} by
$B/\mu$-type terms.

 The eigenvalues  of this matrix are squared masses of physical the CP-even neutral Higgses at tree level, denoted
  $\lambda=m^2_{H^0}$. They  must
  satisfy  the equation  $\det\left( M^2_{6S}-\lambda ~I_6\right)=0$, or equivalently
 \bea \lambda \left[\lambda-
 \left(1+t^2_{\beta}\right)\left(\frac{b_{\chi}}{t_{\beta}}+\frac{g^2}{4}(u^{\prime2}+w^{\prime2})
 \right)\right]f(\lambda)=0 \label{det1}\\
  \mathrm{with}  \;  \;\;
 f(\lambda)=a \lambda^4 + b
 \lambda^3+c\lambda^2+d\lambda+e.
 \label{ptdinhthuc1}\eea
Equation (\ref{det1}) has one massless solution and one exact
massive solution $\lambda=m^2_{A_3}$. The massless Higgs is eaten
by $X$ boson. The function $f(\lambda)$ can be reduced  to a simpler form by
 defining a new variable  as follows:
\be \lambda= X \times m^2_X.\label{factor1}\ee
 From (\ref{linear8}) and  (\ref{notation2}) we get
\bea
 b_{\chi}&=& \frac{1}{2} m^2_{A_2}s_{2\beta}=\frac{1}{2}m^2_X
 k_1 s_{2\beta}, \hs
 b_{\rho}= \frac{1}{2} m^2_{A_1}s_{2\gamma}= \frac{1}{2}m^2_X k_2s_{2\gamma} ,\crn
 g'&=& t \times g \label{factor2}\eea
 with $$t^2=\frac{18s^2_W}{3-4s^2_W}.$$

We define  the quantity
\be \epsilon=
\frac{m^2_W}{m^2_X}=\frac{v^2+v^{\prime2}}{(u^{\prime2}
+w^{\prime2})(1+t^2_{\beta})},\label{epsilon1}\ee which measures
ratio of two spontaneous breaking scales $\mathrm{SU(2)}_L$ and
$\mathrm{SU(3)}_L$. Based on the calculation in
\cite{susyec,haihiggs} we get a relation
\be \epsilon \simeq \frac{m^2_W}{m^2_{Z'}}\times
\frac{4c^2_W}{4c^2_W-1}, \label{epsilon2}\ee
where $m_{Z'}$ is the mass of heavy neutral Hermitian boson $Z'$
and $\theta_W$ is the Weinberg angle,  $c_W=\cos\theta_W$. The
current bound of $m_{Z'}$ is $m_{Z'}>2500$ GeV \cite{coutinho}
leading to the result $\epsilon<2.0 \times 10^{-3}$ which can be
used to find solutions of Eq. (\ref{ptdinhthuc1}) approximately.

The equation  $f(\lambda)=0$ now can be written in the form of \be
g(X)= A X^4+BX^3+ C X^2+ D X + E=0,\label{newfla1} \ee
 where
 \bea  A&=&1,\crn
  B&=& -\left(4c^2_W h^2_W+k_1+k_2+4h^2_W\times\epsilon\right),\crn
 C&=&4c^2_Wh^2_W\left(k_1+k_2c^2_{2\beta}\right)+k_1k_2
 +h^2_W\left(1+k_1c^2_{2\gamma}+k_2\right)\times
 \epsilon,\crn
 D&=&
 -4c^2_Wh^2_Wk_1k_2c^2_{2\beta}-4h^2_W\left[k_2c^2_{2\beta}
 +c^2_{2\gamma}k_1\left(1+k_2\right)\right]
 \times\epsilon,\crn
  E&=& 4h^2_Wk_1k_2c^2_{2\gamma}c^2_{2\beta} \times\epsilon.
 \label{factor1}\eea
The function $g(X)$ will be used to  estimate the approximate mass
eigenvalues of real neutral Higgses in the following section. We
will study
 in more detail  the mass spectrum of neutral Higgs with some
assumptions on the soft parameters. Now let us consider
 the charged Higgs mass spectrum.

\subsection{\label{cHiggs2}Charged Higgs}

In the basis of $(\chi^{+},\ \chi^{+\prime},\ \rho^{+}_1,\
\rho^{+}_2,\ \rho^{+\prime}_1,\ \rho^{+\prime}_2)$, the squared
mass matrix can be written as
 \bea
 M^2_{6\mathrm{charged}}=\frac{g^2}{4} \left(
 \begin{array}{cccccc}
 m^2_{\chi^-\chi^{+}} & m^2_{\chi^-\chi^{\prime+}}& uv &
vw & -uv^\prime & -v^\prime w \\
&  m^2_{\chi^{\prime-}\chi^{\prime+}} & -vu^\prime
 & -w^\prime v & v^\prime u^\prime & v^\prime w^\prime \\
 &  & m^2_{\rho^-_1\rho^+_1} &m^2_{\rho^-_1\rho_2^+} & -\frac{4b_{\rho}}{g^2}-vv' &  0 \\
  &  &
 & -m^2_{\rho^-_2\rho_2^+}&0 &-\frac{4b_{\rho}}{g^2}-vv'  \\
  &  &
 &  & m^2_{\rho_1^{-\prime}\rho_1^{+\prime}} &
 m^2_{\rho^{-\prime}_1\rho^{+\prime}_2}\\
  &  &  &  &  &
 m^2_{\rho_2^{-\prime}\rho_2^{+\prime}} \\
 \end{array}
 \right).\crn \label{chargeh}
 \eea
 Detailed formulas for  elements of the matrix are shown in Appendix
\ref{chiggsmass}.

The masses of charged Higgses are solutions of the Eq.
$\mathrm{Det}(M^2_{6\mathrm{charged}}-\lambda I_6 )=0$. Each
solution $\lambda=m^2_{H^{\pm}}$ corresponds to one mass
eigenvalue of $M^2_{6\mathrm{charged}}$ and $I_6$ is the $6\times
6$ unit matrix. Changing variables  as in the case of the neutral
Higgses, we obtain the equation
 \be
 X^2\left[\lambda-\left(m^2_{A_1}+m_W^2\right)\right]\times
 f(X)=0
 \label{MChargedH1}\ee
with $X= m^2_{H^{\pm}}/m^2_X$ and where $m_W$ is the mass of the $W$
boson. The function $f(X)$ is a polynomial of degree
 3,  presented as
\bea  f(X)&=& X^3+AX^2+ BX +C, \hs \mathrm{where}\label{echargedH}\\
  A&=& -(1+k_1+k_2+\epsilon) ,\crn
 B&=&
 -c^2_{2\beta}+k_1(1+c_{2\gamma}c_{2\beta}+k_2)+\left[k_2+c_{2\gamma}c_{2\beta}(2+k_2)\right]\times
 \epsilon-c^2_{\gamma}\times \epsilon^2,\crn
 C&=&(1+\epsilon)\left[c_{2\beta}-c_{2\gamma}(\epsilon+k_1)\right]
 \left[c_{2\beta}(1+k_2)-c_{2\gamma}\epsilon\right].\label{3chargedHigss1}\eea
 For the charged Higgs sector, there are two Goldstone bosons
 eaten by the $W^{\pm}$ and $Y^{\pm}$ bosons. There is an exact value
 of the mass, $m^2_{H^{\pm}_4}= m^2_W+m^2_{A_1}$. The three other values
 will be investigated in the following section.

\section{\label{constrainH} Constraint to  Higgs  masses}

 As stated above, in this section we  will investigate
  in  more detail th mass spectrum of the Higgses. We will see that there exist many relations among Higgs
 masses, soft parameters and $\mu_{\rho,\chi}$ terms in the
 scalar potential. First,  from (\ref{linear8}),
 (\ref{condition1}), (\ref{masscHiggs}) and the lower
  constraint to the CP-odd neutral Higgs masses from a recent experiment, we conclude that all
 parameters of the model must  be  above  the electroweak
 breaking scale.
 Furthermore, the equations in (\ref{condition1}) indicate that the
 soft-breaking parameters must be smaller than the $\mathrm{SU(3)}_L$
 breaking scale, and  $c_{2\gamma}$ should not be too small.

 To continue, we  will investigate masses of neutral and charged Higgses
 in two cases listed in (\ref{case1}) and  (\ref{case2}). From
 these two cases and (\ref{masscHiggs}),  it is easy to
prove that  $m^2_{A_1}$ and $m^2_{A_2}$ have the same order as
parameters in (\ref{case1}). We will concentrate on the
 values  of  $m^2_{A_1}$ and $m^2_{A_2}$ in the following sections.

 \subsection{Case1: Soft-parameters in the electroweak breaking scale}

This case is expressed in  (\ref{case1}). The result is
that $k_1,k_2$, and $c_{2\beta}$ are of the order
$\mathcal{O}(\epsilon)$. So we define
\be  k_1= k'_1\times \epsilon,\hs  k_2= k'_2\times
\epsilon\label{para1}\ee
 with $k'_1,~k'_2 \sim \mathcal{O}(1)$. The factor $c_{2\beta}$ will be considered later.
  Based on the Viet theorem, the equations given in
 (\ref{factor1}) show that the Eq. (\ref{newfla1}) produce  four positive solutions related to physical
 squared masses of Higgses.
  Without loss of generality, we denote
 these four solutions  as $X_1 \leq X_2 \leq X_3 \leq  X_4$.
 The Viet theorem gives the four solutions satisfying the conditions
\bea
  \sum_{i=1}^4X_i&=& 4c^2_W h^2_W+k_1+k_2+4h^2_W\times\epsilon,\crn
 \sum_{i<j;i,j=1}^4X_iX_j &=&4c^2_Wh^2_W\left(k_1+k_2c^2_{2\beta}\right)+k_1k_2+h^2_W
 \left(1+k_1c^2_{2\gamma}+k_2\right)\times
 \epsilon,\crn
 \sum_{i<j<k}X_iX_jX_k&=&
 4c^2_Wh^2_Wk_1k_2c^2_{2\beta}+4h^2_W\left[k_2c^2_{2\beta}+c^2_{2\gamma}k_1\left(1+k_2\right)\right]
 \times\epsilon,\crn
  X_1X_2X_3X_4&=& 4h^2_Wk_1k_2c^2_{2\gamma}c^2_{2\beta} \times\epsilon.
 \label{viet1}\eea

 Because of the existence of the $4c^2_W h^2_W$ term in the first equation
 of (\ref{viet1}), there must be at least one heavy Higgs which is
 equivalent to $\mathcal{O}(m^2_X)$. The fourth equation shows that
 $X_1X_2 X_3X_4 \leq  \mathcal{O}(\epsilon)$ in this case.  So there is at
 least one light Higgs having mass related with $X_i\leq \mathcal{O}(\epsilon)$.
 We first estimate this mass  by  assigning
 $X_1=X'_1\times\epsilon$ with $X'_1\leq \mathcal{O}(1)$.
 Inserting this $X_1$
  into Eq. (\ref{newfla1}) then setting the
 factor of the lowest order of $\epsilon$ to vanish, we have
 \be  (X'_1-k'_1c^2_{2\beta})\left[c^2_W X^{\prime2}_1 -(1+c^2_W
 k'_1)X'_1 +c^2_{2\gamma} k'_1\right] =0.\label{lihgtnhiggs1}\ee
The equation (\ref{lihgtnhiggs1}) indicates that there are three light
Higgses. But one  of them relates with $X'_1$ such that
\be  X'_1= k'_1c^2_{2\beta} \sim\frac{m^2_{A_1}}{m^2_W}\times
|c_{2\beta}|^2. \label{lightestnHiggs}\ee
 This value is too small because the factor
 $c^2_{2\beta}\sim\mathcal{O}(\ep^2)$ given in
Eq. (\ref{condition1}) and the soft parameter scale is the same
as that of $\mathrm{SU(2)}_L$ breaking. It is then
 $$ X'_1=\frac{m^2_{H^0_1}}{m^2_W}\sim
 \frac{m^2_{A_1}}{m^2_W}\times\left(\frac{m^2_W}{m^2_X}\right)^2\hs
 \rightarrow m_{H^0_1}\sim m_{A_1}\times \mathcal{O}(10^{-3}).$$
 Because $m_{A_1} \sim \mathcal{O}(m_W)$ in this case, if this is
 SM  Higgs $m_{H^0_1}$ is too small compared with recent
 experimental bound from LEP \cite{Barat1}. If not, one of two
 remaining solutions in (\ref{lihgtnhiggs1}) will be identified with
 the value around  125.5 GeV. The formula presenting these two
 values are
 \be
  m^2_{H^0_{2,3}}\simeq \frac{1}{2}\left(m^2_Z+m^2_{A_1}\mp\sqrt{\left(m^2_{A_1}-m^2_Z\right)^2
   + 4s^2_{2\gamma}m^2_Zm^2_{A_1}}\right).
  \label{lnHiggses1}\ee
Formula (\ref{lnHiggses1}) is of exactly the same form as that
presented for neutral Higgs masses in the  MSSM. From previous
work for the MSSM we immediately obtain some interesting
consequences. At
  tree level the lighter Higgs gets  a mass
which is smaller than of $m_Z|c_{2\gamma}|$. This Higgs is
normally  identified with the like-SM  Higgs discovered at the LHC
\cite{higgsdicovery1,higgsdicovery2} because its mass can increase
after including loop corrections. On the other hand, some recent
works also were concerned with a case named "\emph{low}-$M_{H}$ scenario"
where the heavier Higgs corresponds to the discovered state
\cite{lowmH}. Although  this case predicts light charged Higgses,
the parameter space is very small. This is because it requires all
of these light Higgses to have heavily suppressed couplings to the
gauge bosons to escape the search of LEP.

From the above investigation,  the SUSYE331 soft parameters considered
at the $\mathrm{SU(2)}_L$ symmetry breaking are not the favorite
choice. They should be in the $\mathrm{SU(3)}_L$ breaking scale.
 It is  case 2 that we concentrate on in
this work.
\subsection{Case2: Soft-parameters in the $\mathrm{SU(3)}_L$ breaking scale}
\subsubsection{CP-even neutral Higgses}

 The Higgs sector in this case is very complicated.
 Mathematically, exact solutions of  the polynomial equations  (\ref{newfla1}) and
 (\ref{echargedH}) can be determined, but  they are too long; also  it is very hard to see any  physics in these expressions.
 Instead, we firstly find approximate solutions of the mass
  eigenvalues based on the very small values of $\epsilon$.

  For  light neutral Higgses, the last equation in (\ref{factor1}) shows that there
  is only one light neutral Higgs.
  Being of the order of
  $\mathcal{O}(\epsilon)$, the squared mass of  this Higgs is given as $X_1= X'_1\times
  \epsilon+ \mathcal{O}(\ep^2)$ where $X'_1\sim \mathcal{O}(1)$. Inserting this value
  into (\ref{newfla1}) then forcing the factor of
  the  lowest order of
  $\epsilon$ to be zero, we have
  \be  X_1'\simeq \frac{c^2_{2\gamma}}{c^2_W}\hs \mathrm{equivalent}
  \hs m^2_{H^0_1}\simeq M^2_{Z}c_{2\gamma}^2.  \label{nlHiggs1}\ee
 This formula for  neutral Higgs mass is
  completely  the same as that in the
 case of the MSSM.
 Furthermore,  the contribution
 from the next leading order
  is proportional to $(\frac{1}{2}m_W\times\ep)\sim 0.08$
 GeV. So the mass of the light Higgs needs to get major corrections from the
 loop contributions.

   For the three heavy neutral Higgses, we denote their masses as $X_i=X_i'+ X''_i\times
   \epsilon$ where  both $X'_i,~X''_i\sim \mathcal{O}(1)$ and $i=2,3,4$. Then  these
   masses can be written in the
   form
   \be  m^2_{H^0_i}=
  X_i' m^2_X + X''_i\times m^2_W+ \mathcal{O}(\epsilon)\times m^2_W. \label{n0hvh1} \ee
 The main contributions to the heavy Higgs masses
  come from  $X_i\times m^2_X \sim m^2_{H^0_i}$, namely
 \bea m^2_{H^0_2}\simeq  X'_2 m^2_X&=& m^2_{A_1}, \label{mheavyH1}\\
   m^2_{H^0_{3,4}}\simeq  X'_{3,4}\times m^2_X&=& \frac{1}{2}
   \left( m^2_{A_2}+m^2_{Z'}\mp \sqrt{\left( m^2_{A_2}-m^2_{Z'}\right)^2
  +4m^2_{Z'} m^2_{A_2}s^2_{2\beta}}\right),\crn\label{mheavyH23}\eea
   where $m_{Z'}$ is the mass of the neutral $Z'$ boson \cite{Dong3},
   $ m^2_{Z'}= 4 m_X^2c^2_W/(4c^2_W-1)$.
  The values of $X''_i$ are computed from $X'_i$ based on the following
 formula:
\bea X''_i&=&\frac{A_0}{B_0}\hs \mathrm{where}\crn
A_0 &=&
4h^2_W\left(X'_i-k_1c^2_{2\gamma}\right)\left(X^{\prime2}_i-(k_2+1)X'_i+k_2c^2_{2\beta}\right),\crn
B_0&=&
4c^2_Wk_2(2X'_i-k_1)c^2_{2\beta}+X'_i\left[4h^2_Wc^2_W\left(2k_1-3X'_i\right)\right.\crn
&+&\left.2k_1k_2-3(k_1+k_2)X'_i+4X_i^{\prime2}\right].
\label{xpp1}\eea
It is noted that $X''_i$ is the correction to the squared Higgs
masses. For the correction of Higgs masses, using Eq.
(\ref{mheavyH1}), we can get  approximate values of the Higgs masses:
\bea m_{H^0_i}&=&\sqrt{ X'_i m^2_X + X''_i\times m^2_W+
\mathcal{O}(\epsilon)\times m^2_W}\crn&\simeq& m_X\times
\sqrt{X'_i}+\frac{X''_i}{\sqrt{X'_i}}\times \frac{m_W^2}{m_X}.
\eea
If we assume that the scale $m_X\simeq \mathcal{O}(\mathrm{TeV})$,
the correction to the Higgs mass at the next leading order is
$X''_i/\sqrt{X'_i}\times 2.4$ GeV. This correction is too small
compared with heavy Higgs mass of TeV scale. So, in our
calculation, this correction  can be ignored.   For more details,
the analytic formulas of neutral Higgs masses can be found in
Appendix \ref{ehiggsmaas}.

\begin{figure}[h]
  \centering
\epsfig{file=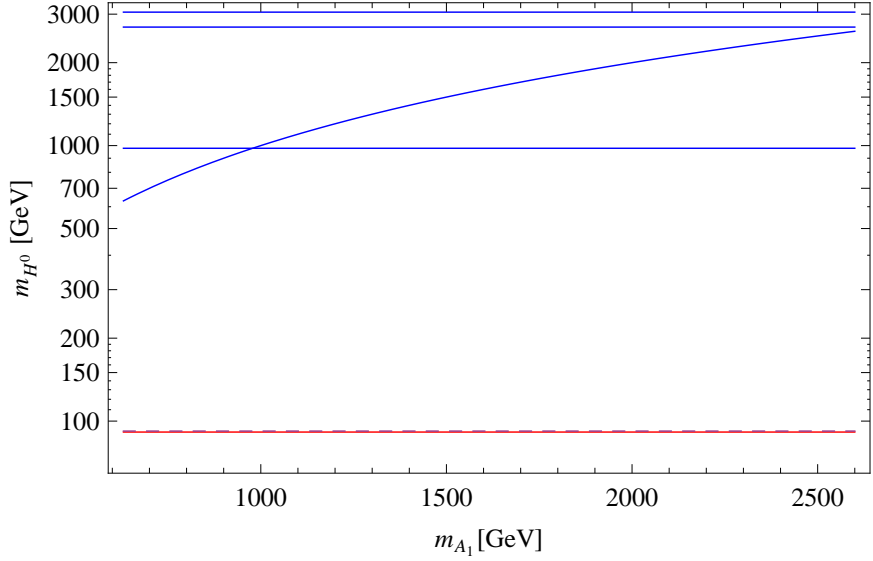,width=0.75\linewidth,clip=}
  \caption{ Plots of $m_{H^0_j}$ ($j=1,2,...,5$) as functions of $m_{A_1}$.
  The parameters are fixed as $m_X=2.5$ TeV, $m_{A_2}=1.0$ TeV,
  $\frac{u^2+u^{\prime2}}{v^2+v^{\prime2}}=10^{-4}$ and
  $m_W=80.4$ GeV,  $t_\gamma=50$, $t_\beta=10$. The red line presents mass of
   the lightest neutral Higgs. The dashed line fixes the values of
   $m_Z\simeq 92.0$ GeV.}\label{nHmass1}
\end{figure}

For illustration of our results, all analytic formulas of the neutral
Higgs masses can be compared with the numerical investigation
 shown in fig.\ref{nHmass1}. In this figure, we use
  Mathematica 7.0  directly to find the eigenvalues of the squared mass matrix (\ref{higgstrunghoa}).
  It is easy to see that the four blue
 curves represent four heavy Higgs masses, while the lightest Higgs
 has mass $m_{H^0_1}\simeq m_Z$ when $t_\gamma\gg 1$. All of
 these masses are consistent with those shown by our analytic
 results. This will be helpful to estimate the mass eigenstates of
 these Higgses in the Appendix  \ref{ehiggsmaas}.

In conclusion, the SUSYE331 model has  five physical CP-even
neutral Higgses, including one light Higgs and four other heavy
Higgses. The light Higgs can be identified to the Standard-Model-like Higgs. One of the heavy Higgses has exactly a mass $m_{H^0_5}$
at the tree level which obeys $m^2_{H^0_5}= m^2_{A_2}+ m^2_{X}$. The squared
masses of the three other  Higgses can be approximately computed up to
$\mathcal{O}(\epsilon)\times
 m^2_W$. The above analysis makes some interesting
 properties of the SUSYE331 clear. Although  the model
 has four
 Higgs multiplets, they separate into two pairs having different absolute  $\mathrm{U(1)}_X$
 charges. Two Higgses in each pair have opposite signs in order to
 cancel the gauge anomaly.  The appearance of the
 Higgses in pairs  makes
  the SUSYE331  have  many similar properties to
 the  MSSM. In particular, while  $\rho$  and $\rho'$ couple with all
 leptons and quarks, $\chi$  and $\chi'$ do not couple with the leptons.
 So $\rho$ and $\rho'$ play  the same role as Higgses in the MSSM.
 Furthermore, if the CP-odd neutral  Higgs is very heavy, the light
CP-even neutral Higgs,  mass  at the tree level  has
  exactly  the form as in (\ref{nlHiggs1}) where
  $t_\gamma$ in  the SUSYE331
   plays the same role as $t_\beta$ in  the  MSSM.
  This value is smaller than the mass of the $Z$ boson, $m_Z\simeq
  92$ GeV.  Compared with the 125.5 GeV value of the Higgs  mass discovered recently in
  the  LHC,  the MSSM needs  large values of $|c_{2\beta}|$.
   $t_\beta$  should also be large, corresponding to large
  corrections from the  squark loops for the Higgs mass in order to  get a consistent light Higgs mass.
  The case of SUSYE331 is a bit different. Apart from
  $t_\gamma$ there appears a new parameter $t_\beta$ defined as
  the ratio of $w$ and $w'$ which are two VEVs of $\chi$ and
  $\chi'$. One can see that the light Higgs state
  is a mixing of all neutral components of the four Higgs
  multiplets. As a result, corrections to this Higgs mass will
  come from  squark loops related with both $t_\gamma$ and
  $t_\beta$.

  The mass of the lightest Higgs will easily and naturally reach the
  value of a recent experimental result if loop corrections are included. This can be realized through the
 well-known results calculated for the  MSSM \cite{Martin,ellis,corr}, where
the largest one-loop   corrections to $m^2_h$ arise  from the top quark and stop scalar.
    In the SUSYE331 model,  choosing a simplifying   case based on
\cite{ellis}  we can show that the lightest Higgs mass can get a
contribution of a one-loop correction similar to those of the  MSSM. The
details are presented in the appendix \ref{correctionNH} and the
figure \ref{masslH1} presents the mass of the lightest Higgs according
to (\ref{lightm3}). In a more accurate calculation,
the mixing between left and right stops should be included; then
the case will be the same as that called the decoupling limit,
indicated in \cite{Martin} (section 7). Apart from this, we
believe that one-loop corrections from the very heavy exotic
quarks and their superpartners may also increase the mass of this
lightest neutral Higgs. This topic is out of the scope of this
work. The simple estimation in this work is an illustration enough
to show that the CP-even neutral Higgs spectrum of the SUSYE331 is
consistent with present experimental  results. Because $t_\gamma$
is larger than 1 ($\frac{\pi}{4}<\gamma<\frac{\pi}{2}$)
 we get the constraint $c_{2\gamma}<0$.
\begin{figure}[h]
  \centering
\epsfig{file=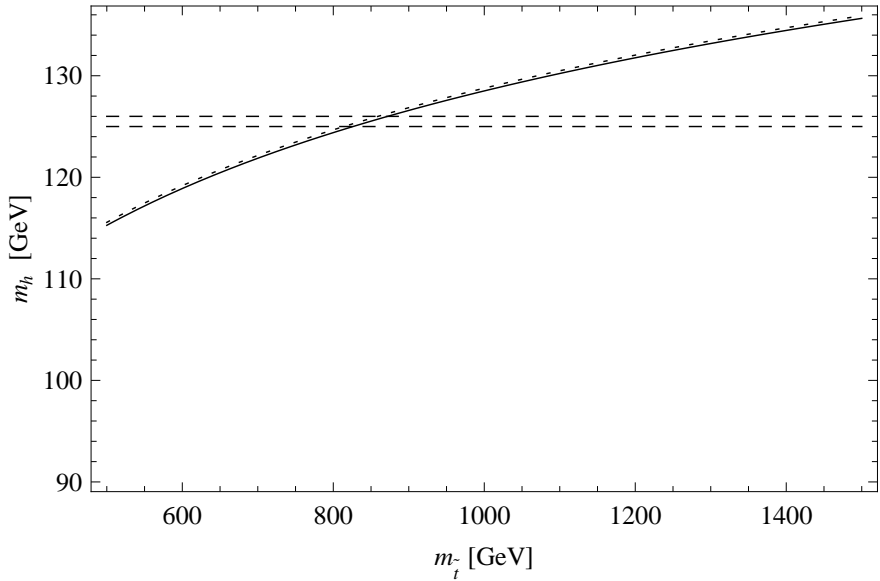,width=0.75\linewidth,clip=}
  \caption{ Mass of the lightest CP-even neutral Higgs including one-loop correction of the top and stop
 quark. The black (dotted) curves present the mass in the SUSYE331 (MSSM) as function of stop
 quark.Two dashed line correspond to 125 and 126 GeV. In the case of the SUSYE331
  $m_X=2.0$  TeV is chosen}\label{masslH1}
\end{figure}

   One more comment needs to be added here. At this scale of soft
 parameters, light Higgs $m^2_{H^0_1}$ in Eq. (\ref{nlHiggs1}) and a
 heavy Higgs in Eq. (\ref{n0hvh1}) contain many similar properties to those in
 the  MSSM,  while other Higgses  are characterized for $\mathrm{SU(3)}_L$ scale. So we can use
 many known properties of the MSSM to study these like-MSSM  Higgses. Also, the
 CP-odd neutral Higgs $H_{A_1}$ in (\ref{pmase1}) carries
 properties of that in the MSSM. As we will show in the next section, the
  Higgs sector in  the SUSYE331 is separated into
   two parts. The first part is closely related to MSSMs while
  the second is related to  $\mathrm{SU(3)}_L\times\mathrm{ U(1)}_X$
  properties.
\subsubsection{Charged Higgs}

If all soft parameters live on the $\mathrm{SU(3)}_L$ scale, the
second formula given in  (\ref{condition1}) shows that the values of
$c_{2\beta}$ should not be too small. Applying this constraint to
the  Eq. (\ref{3chargedHigss1}), one can prove that all solutions
of (\ref{MChargedH1}) correspond to very large values of
charged Higgs masses. Similar to the case of neutral Higgs, if  we
denote $X_i=X'_i+ X''_i\times \epsilon$ ($i=1,2,3$)then
\be  m^2_{H_i^{\pm}}= X_i\times m^2_X= X'_i\times m^2_X + X''_i
\times m^2_W+\mathcal{O}(\epsilon)\times m^2_W, \label{cHmass1}\ee
 where the  main contributions to the three charged Higgs masses are
\bea  m^2_{H^{\pm}_1}&\simeq&  X'_1\times m^2_X= m^2_X+ m^2_{A_2},\label{chiggs1}\\
 m^2_{H^{\pm}_{2,3}}&\simeq& X'_{2,3}\times m^2_X=\frac{1}{2}\left( m^2_{A_1} \mp \sqrt{
\left(m^2_{A_1}-2m^2_Xc_{2\beta}c_{2\gamma}\right)^2+4m^4_Xc^2_{2\beta}s^2_{2\gamma}}\right),\crn
\label{cHmass2}\eea
 and $X''_i\equiv a_x/b_x$  depends on $X'_i$ according to the following
 formula:
 \bea a_x&=&-c_{2\beta}c_{2\gamma}
 \left[ 1+(k_1+1)(k_2+1)+(k_2+2)X'_i\right]\crn&+&
  c^2_{2\gamma}k_1+c^2_{2\beta}(1+k_2)+k_2X'_i-X_i^{\prime2},
 \crn
 b_{x}&=& c^2_{2\beta}-c_{2\beta}c_{2\gamma} k_1-k_1(k_1+k_2)+2(1+k_1+k_2)X'_i-3X_i^{\prime2}.
 \label{xppc}\eea
 We need to emphasize that masses of Higgses in (\ref{cHmass2}) must
 be positive. It  corresponds to the  condition:
 \be c_{2\beta}\left( c_{2\beta}-k_1c_{2\gamma}\right) <0.\label{positivemas} \ee
If  so  then $ k_1 c_{2\gamma}<c_{2\beta}<0$ because
$c_{2\gamma}<0$.  From this we have $\pi/4<\beta<\pi/2$ and
$t_\beta>1$.

There is another way to deduce an exact constraint,  which is
stricter than the constraint given in Eq. (\ref{positivemas}). By
applying the Viet theorem to Eq. (\ref{3chargedHigss1}) with
three charged Higgs masses $X_1,~X_2$ and $X_3$, we have
$X_1X_2X_3=-(1+\epsilon)\left[c_{2\beta}-c_{2\gamma}(\epsilon+k_1)\right]$
$\times \left[c_{2\beta}(1+k_2)-c_{2\gamma}\epsilon\right]>0$. In
case of
  $\epsilon\ll 1$  it leads to  a
   consequence that $ (c_{2\beta}-c_{2\gamma}k_1)c_{2\beta}(1+k_2)<0$, the
 same result as shown in Eq. (\ref{positivemas}).  Combining with the condition of
 $c_{2\gamma}<0$, we get an exact condition for positivity of all charged Higgs masses:
 $(k_1+\epsilon) c_{2\gamma}<c_{2\beta}<
\frac{c_{2\gamma}\epsilon}{1+k_2}<0$,
 which implies that
\be \frac{(m_{A_1}^2+m_W^2)c_{2\gamma}}{m_X^2}<c_{2\beta}
<\frac{c_{2\gamma}m^2_W}{m_X^2+m^2_{A_2}}<0.
 \label{positivemasco}\ee
 If this condition is satisfied, then all charged Higgs masses in
 the SUSYE331  are of the order of $\mathrm{SU(3)}_L$ scale. Of course,  on this
 scale, there is no massless charged  Higgs in this model and
 all of these masses  are much larger than current experimental bound  at LEP
 \cite{abb}.
\begin{figure}[h]
  \centering
\begin{tabular}{cc}
\epsfig{file=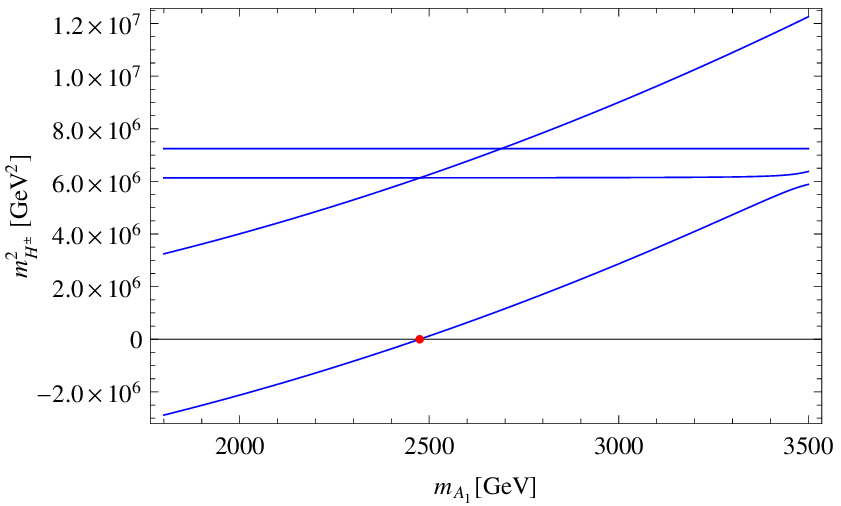,width=0.495\linewidth,clip=} &
\epsfig{file=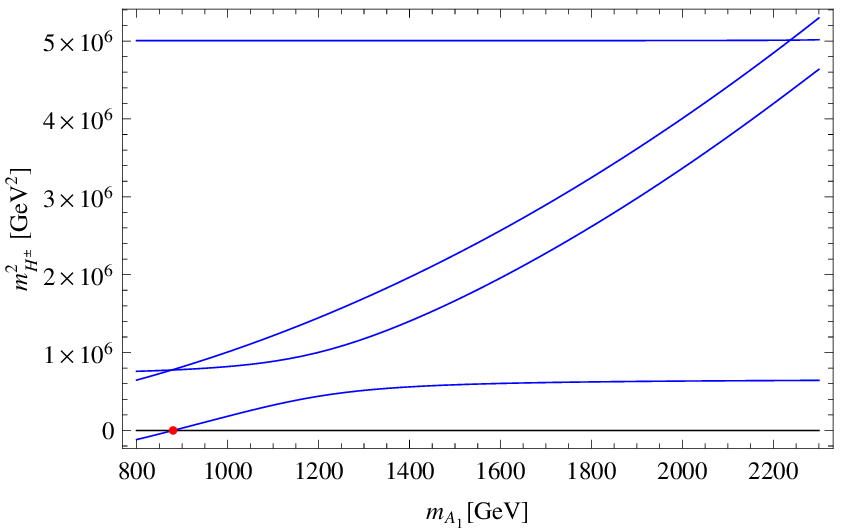,width=0.495\linewidth,clip=}
\\
\end{tabular}
  \caption{ Plots of $m^2_{H^{\pm}_i}$ as functions of $m_{A_1}$.
  The parameters are fixed as $m_X=2.5$ TeV (left panel)
   and $m_X=2.0$ TeV (right panel), $m_{A_2}=1.0$ TeV, $\frac{u^2+u^{\prime2}}{v^2+v^{\prime2}}=10^{-4}$ and
  $m_W=80.4$ GeV.
  The left panel corresponds to large values of $t_\gamma$ and
  $t_\beta$: $t_\gamma=50.$, $t_\beta=10$. The right panel
  corresponds to smaller values of $t_\gamma$ and
  $t_\beta$: $t_\gamma=5.0$, $t_\beta=1.2$. The red points
  imply the values of $m^2_{A_1}=\frac{m^2_X
  c_{2\beta}}{c_{2\gamma}}-m^2_W$ giving squared mass of lightest charged Higgs
   $m^2_{H^{\pm}_2}\simeq0$.}\label{cHiggsmass1}
\end{figure}

\begin{figure}[h]
  \centering
\begin{tabular}{cc}
\epsfig{file=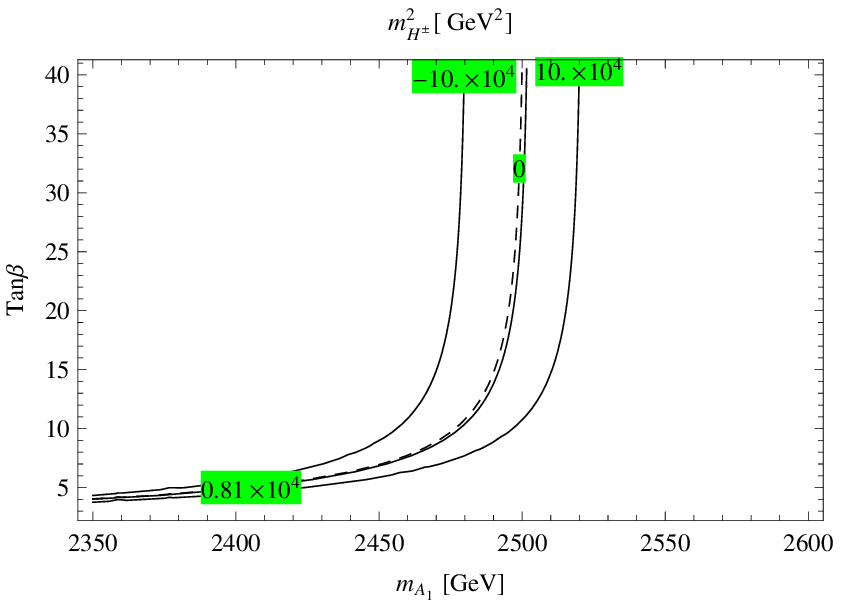,width=0.495\linewidth,clip=}
&
\epsfig{file=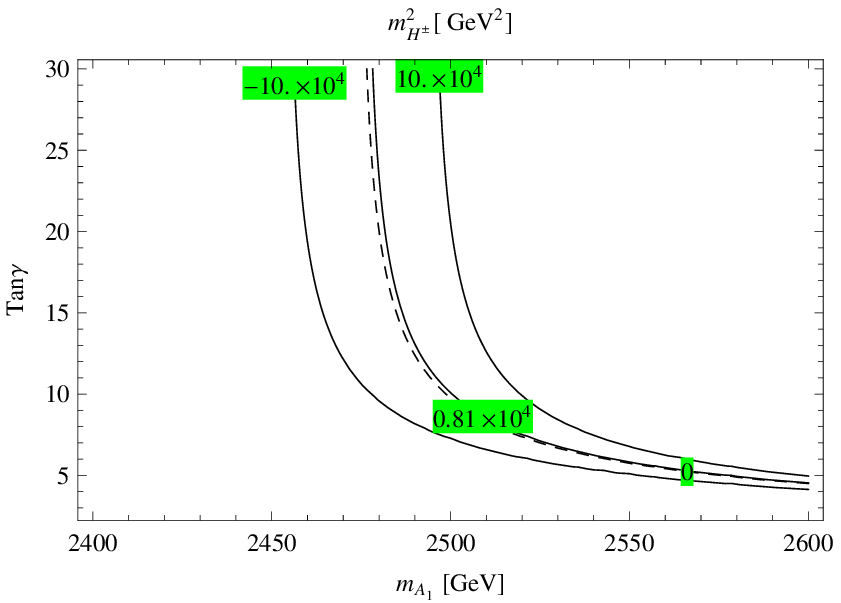,width=0.495\linewidth,clip=}
\\
\end{tabular}
\caption{ Contours of lightest values of $m^2_{H^{\pm}}$ as
functions of two variables : $(m_{A_1},~t_\beta)$ (left panel) or
$(m_{A_1},~t_\gamma)$ (right panel).
  The parameters are fixed as $m_X=2.5$ TeV, $m_{A_2}=1.0$ TeV,
  $\frac{u^2+u^{\prime2}}{v^2+v^{\prime2}}=10^{-4}$  and
  $m^2_W=80.2$. In addition $t_\gamma= 30$ for the left panel and $t_\beta=10$ in the right
  panel. The dashed line corresponds to $m^2_{H^{\pm}}=0$.}\label{cHiggsmass2}
\end{figure}

Finally, as an illustration for our qualitative  estimations  we
will numerically investigate some  cases of charged Higgs masses.
The results are shown in Figs. \ref{cHiggsmass1} and
\ref{cHiggsmass2}. The left panel of Fig. \ref{cHiggsmass1} shows
the case of large $t_\gamma$ and $t_\beta$ where we can fix
$c_{2\gamma}\simeq c_{2\beta}=-1$. Inserting these values into
(\ref{cHmass2}) we have two values $m^2_{H^{\pm}}=\{
m^2_{X},~m^2_{A_1}-m^2_X\}$. This means that in order to cancel
tachyon Higgs $m_{A_1}$ must be larger than $m_X$. A strict
constraint of $m_{A_1}$ comes from (\ref{positivemasco}):
$m^2_{A_1}>\left|\frac{c_{2\beta}}{c_{2\gamma}}\right|m^2_X-m^2_W$.
This limit value of $m_{A_1}$ is represented by the red points in Fig.
\ref{cHiggsmass1}.
It is easy to see that the two constant lines in the left panel
represent two  values $m^2_{H^{\pm}}=\{ m^2_{X}+m^2_{A_2},~m_X^2\}$
while the two other curves show values of $m^2_{H^{\pm}}=\{
m^2_{A_1}+m_W^2,~m^2_{A_1}-m_X^2\}$. These two curves are parallel
because they are different from each other at constant values
$m^2_X+m^2_W$. This property does not occur in the case of small
$t_\gamma$, as shown in the right panel of Fig. \ref{cHiggsmass1}. In
all cases, there always exists a lower constraint of $m_{A_1}$ to
cancel the tachyon charged Higgs. This value lies at
$\mathrm{SU(3)}_L$ scale unless $|c_{2\beta}|$ ($t_{\beta}$) is
small, as we illustrate in the left panel of Fig.
\ref{cHiggsmass2}. This also shows the consequence that the
SUSYE331 still contains a light charged Higgs if  the value of
$m^2_{A_1}$ is very close to the values of
$\left|\frac{c_{2\beta}}{c_{2\gamma}}\right|m^2_X-m^2_W$.

There is an interesting consistence of the model that can be
 seen in Fig. \ref{cHiggsmass2}. It shows the contours of the
 lightest mass of the charged Higgs
 $m^2_{H^{\pm}}$ as functions of $m^2_{A_1}$ and  $t_\beta$
 ($t_\gamma$). The allowed regions correspond to the condition
  $m^2_{H^{\pm}}>90^2$ [GeV] at tree
 level.  As we have discussed, the
 model  requires a large $t_\gamma$ to get the consistent lightest neutral Higgs.
 Fortunately, the allowed  region on the right panel favors both large
 $t_\gamma$ and $m_{A_1}$. The small values of $t_\gamma$
 require very large values of $m_{A_1}$. On the other hand, the
 allowed region  with large $m_{A_1}$ in the left panel also supports large values of
 $t_\beta$.  We can see that in the
 limit of large $t_\gamma$ ($t_\beta$)   lightest charged
 Higgs mass almost does not depend on the values of $t_\gamma$
 ($t_\beta$) while it is very sensitive to the variance of
 $m_{A_1}$.
\section{\label{mssmvs331}MSSM  Higgses vs. SUSYE331 Higgses}

To compare more precisely the properties of the  MSSM Higgs
spectrum with some Higgses in the model under consideration we
will investigate the couplings of the Higgs particles. In this
part, we concentrate on the couplings of Higgses in the SUSYE331.

Let us briefly review the Higgs spectrum in the MSSM. In
this model, in order to provide  mass for up and down fermions as
well as to  cancel the anomaly, two doublet Higgses $H_u,H_d$ are
introduced. After the symmetry breaking  $SU(2)_L \times U(1)_Y
\rightarrow U(1)_Q$, the gauge bosons $W^\pm, Z$ become massive
particles and the physical Higgs spectrum contains two CP-even
neutral $H,~ h$, one odd-CP neutral $A$ and  two singly charged
Higgses $H^{\pm}$.

In the
SUSYE331 the electroweak symmetry is broken by VEVs: $u,u^\prime, v,
v^\prime$, where $u,u^\prime$ are the VEVs of the first components
of $\chi,~  \chi^\prime$ and the residual values are the VEVs of
$\rho,~ \rho^\prime$. Because the $u,u^\prime$ carry lepton
number, they break  the lepton number. Hence  they must be  small
and  we can ignore them when we estimate the effect of electroweak
breaking. It means that the main contributions to the mass of the SM
particles are obtained by VEVs of $\rho, \rho^\prime$. In other
words, these two Higgses have the same roles as the two Higgs
doublets
 $ H_u$ and $ H_{d}$ in the MSSM. Therefore, to find the
similarity between the Higgs spectrum in the MSSM and the
SUSYE331, we will concentrate on studying five particular Higgses
of the SUSYE331, $H^0_1, ~H^0_2$,
 $H_{A_1}$ and $H^{\pm}_{4}$,  where all of them are related with $\rho$, $\rho'$ and
 $B_{\rho}$-term.

 Let us consider the couplings of $H^0_1, ~H^0_2$,
 $H_{A_1}$ and $H^{\pm}_{4}$ with the SM fermions and gauge bosons.
In the limit of large $t_\gamma$ and $u,u'=0$ , the soft as well
as
 $\mathrm{SU(3)}_L$ parameters are assumed to be  much larger than
 the $\mathrm{SU(2)}_L$ breaking scale, in the sense that $m^2_{A_{1,2}} \gg m^2_Z$. The
physical states $H^0_1, ~H^0_2$,
 $H_{A_1}$ and $H^{\pm}_{4}$ have the following forms:
\bea  \left(%
\begin{array}{c}
  H_{A_4} \\
  H_{A_1} \\
\end{array}%
\right)= \left(%
\begin{array}{cc}
  c_{\gamma} & -s_{\gamma} \\
  s_{\gamma} & c_{\gamma} \\
\end{array}%
\right) \left(%
\begin{array}{c}
  A_{6} \\
  A_{5} \\
\end{array}%
\right) , \hs ~\left(%
\begin{array}{c}
  H^{\pm}_5 \\
  H^{\pm}_4 \\
\end{array}%
\right)= \left(%
\begin{array}{cc}
  c_{\gamma} & s_{\gamma} \\
  -s_{\gamma} & c_{\gamma} \\
\end{array}%
\right) \left(%
\begin{array}{c}
  \rho^{\prime\pm}_1 \\
  \rho^{\pm}_1 \\
\end{array}%
\right)   \label{Arhoevalue1}\eea
 and
\bea  \left(%
\begin{array}{c}
  H^0_{2} \\
  H^0_{1} \\
\end{array}%
\right)= \left(%
\begin{array}{cc}
  s_{\gamma} & -c_{\gamma} \\
  c_{\gamma} & s_{\gamma} \\
\end{array}%
\right) \left(%
\begin{array}{c}
  S_{6} \\
  S_{5} \\
\end{array}%
\right).\label{Arhoevalue2}\eea The non-zero masses of these
particles are given by
 \bea
m_{A_1}^2=\frac{b_\rho}{s_{2 \gamma}}, \hs m_{H_4^\pm
}=m_{A_1}^2+m^2_W , \hs m^2_{H_2^o}=m_{A_1}^2+ O(m^2_W), \hs
m_{H^o_1}=m_Z^2|c_{2\gamma}|.\eea The other particles are massless
and identified with the Goldstone bosons. Based on the physical
states, we can find the couplings of the Higgses $H^0_1,
~H^0_2$,
 $H_{A_1}$ and $H^{\pm}_{4}$  with the SM particles. The
 couplings of them with the SM gauge bosons are listed  in Table
 \ref{HGcouple1}.

From Eq. (\ref{Arhoevalue1}), it can be realized that
the equivalent role of two parameters $\beta$ and $\gamma$ in the
two models\footnote{In fact, signs of some elements in
transformation matrices in the SUSYE331 may be different from the
MSSM; see for example \cite{Djou,Martin}. This also happens in the two
definitions of \cite{Djou} and \cite{Martin} in the  MSSM. The
reason is the difference in signs of the two definitions :
 (i) the $B/\mu$ term in the Lagrangian ; (ii) the mass eigenstates of the
 Higgses. These mathematical differences  do not affect the final
 physical results.}. The formula (\ref{Arhoevalue2}) shows that the case we
are working in, the SUSYE331,  is similar to that of the decoupling
regime in the  MSSM  where $\alpha \rightarrow \beta-\pi/2$. In
this limit, the couplings of the considered Higgses with the SM
gauge bosons given in Table \ref{HGcouple1} are consistent with
those of the Higgses in the MSSM  shown in \cite{Djou}.
\begin{table}[h]
  \centering
  \begin{tabular}{|c|c|c|c|}
    \hline
    Vertex & factor & Vertex & factor \\
    \hline
    $H^0_1W^{+}W^{-}$ &$igm_Ws_{2\gamma}$ &$H^0_2W^{+}W^{-}$ &$-igm_Wc_{2\gamma}$  \\
    $H_{A_1}H^0_2Z_{\mu}$ &$\frac{-g}{2c_W}(p+p')^{\mu}$ & $H^{+}_4H^{-}_4Z_{\mu}$ & $\frac{-gc_{2W}}{2c_W}
    (p+p')^{\mu}$ \\
    $H^{+}_4H^{-}_4A_{\mu}$ & $-ie
    (p+p')^{\mu}$ & $H^0_2H^{\pm}_4W^{\mp}_{\mu}$ &  $\pm \frac{ig}{2}(p+p')^{\mu}$\\
    $H_{A_1}H^{\pm}_4W^{\mp}_{\mu}$ &  $\frac{g}{2}(p+p')^{\mu}$&&\\
    $H^0_1H^0_1W^{+}_{\mu}W^{-}_{\nu}$&$\frac{ig^2}{2}g^{\mu\nu}$&$H^0_2H^0_2W^{+}_{\mu}W^{-}_{\nu}$&$\frac{ig^2}{2}
    g^{\mu\nu}$\\
   $H_{A_1}H_{A_1}W^{+}_{\mu}W^{-}_{\nu}$&
   $\frac{ig^2}{2}g^{\mu\nu}$&$H^0_1H^0_1Z_{\mu}W_{\nu}$&$\frac{ig^2}{2c^2_W}g^{\mu\nu}$\\
   $H^0_2H^0_2Z_{\mu}W_{\nu}$&$\frac{ig^2}{2c^2_W}g^{\mu\nu}$&
   $H_{A_1}H_{A_1}Z_{\mu}W_{\nu}$&$\frac{ig^2}{2c^2_W}g^{\mu\nu}$\\
   $H^+_4H^-_4Z_{\mu}W_{\nu}$&$\frac{ig^2c^2_{2W}}{2c^2_W}g^{\mu\nu}$&
   $H^+_4H^-_4Z_{\mu}A_{\nu}$&$\frac{iegc_{2W}}{c_W}g^{\mu\nu}$\\
    $H^0_2H^{\pm}_4W^{\mp}_{\mu}Z_{\nu}$& $ \frac{ig^2s^2_W}{2c_W}g^{\mu\nu}$&
    $H_{A_1}H^{\pm}_4W^{\mp}_{\mu}Z_{\nu}$& $ \mp\frac{g^2s^2_W}{2c_W}g^{\mu\nu}$\\
    $H^0_2H^{\pm}_4W^{\mp}_{\mu}A_{\nu}$& $ \frac{-ieg}{2}g^{\mu\nu}$&
    $H_{A_1}H^{\pm}_4W^{\mp}_{\mu}Z_{\nu}$& $ \pm\frac{e g}{2}g^{\mu\nu}$\\
    \hline
  \end{tabular}
  \caption{Higgs-gauge boson couplings}\label{HGcouple1}
\end{table}
The couplings of the considered Higgses in the SUSYE331 with the
fermions are listed in Table \ref{table1}. The results show that
the couplings among these Higgses are the same as those in the
MSSM. Finally,  we will investigate the LFV of Higgses decaying  to
leptons in the SUSYE331 model in the following section.

\section{\label{LFVHiggs}Lepton flavor violating decay of Higgs to muon and tauon}
The LFV decays of neutral Higgses in the SUSYE331 were studied in
\cite{giang} based on the parametrization  of slepton mixing in
\cite{Anna1} and the model constructed in \cite{susyec} without
the presence of $B/\mu$-type terms. In this work, we use the
revised model where these $B/\mu$-type terms are added to
guarantee the stability of vacuum of the model. As  a result, the
mass eigenstates of all Higgses in general are different from
those in \cite{susyec,Dong3}. The Higgs sector becomes  more
complicated and it is not easy to represent analytically masses as
well as mass eigenstates of the real neutral Higgses in terms of
original parameters. In the limit of large $t_\gamma$ we can
use the LFV Lagrangian established in \cite{giang},
 \bea - \mathcal{L}^{FV}_{H\mu\tau}  &\simeq& Y_\tau(\Delta^{\rho}_R \mu^c \tau
 + \Delta^{\rho}_L
 \tau^c\mu ) (\rho^{0*}-t_{\gamma} \rho^{\prime 0})
 + \mathrm{H.c.},\label{FCNC2}\eea
 which is not affected by the diagonalization of the neutral Higgs mass
 matrix. As noted in \cite{giang} we recall that $\rho^0$, $\rho^{\prime0}$ are neutral Higgses which
  generate masses for the lepton after spontaneous breaking:
  $\Delta^{\rho}_R$ and $\Delta^{\rho}_L$ are one-loop
  contributions to the LFV Lagrangian. We emphasize that the presence of the $B/\mu$-type terms in
   the  model under consideration  does  not modify the analytic
  formulas of the effective couplings $\Delta^{\rho}_R, \Delta^{\rho}_L$ given in \cite{giang}.

 Unlike previous version,  one of  many features of the SUSYE331 in this work  is the presence of massive
 pseudo-scalar Higgses. Especially, formulas  in (\ref{pmase1}) and (\ref{pmase2})
  imply that only $H_{A_1}$ can decay to leptons.
   Furthermore, it is easy to prove that
  \be\mathrm{Im} (\rho^{0*}-t_{\gamma} \rho^{\prime 0}) =
    \frac{-i}{\sqrt{2} c_\gamma}\times H_{A_1}. \label{pnHiggs}\ee
 For the real neutral Higgses, we cannot find the
  exact mass eigenvalues or mass eigenstates of all these Higgses. The
  approximate estimation presented above just only helps to
  understand some qualitative aspects of them  and also shows
  that the Higgs sector of the model is consistent with
  recent results of experiments. A detailed analysis to estimate
  the mass eigenstates of the neutral Higgses is presented in the
  Appendix \ref{ehiggsmaas}. In this work, the real parts of $\rho^0$  and
  $\rho^{\prime0}$ can be estimated as $S_5= H^0_1
  s_{\gamma}-H^0_2c_{\gamma}$ and $S_6=H^0_1
  c_{\gamma}+ H^0_2s_{\gamma}$. The effective
  Lagrangian for the LFV decays of neutral Higgses are
  \be \mathcal{L}^{FV}_{H^0\mu\tau} = \frac{Y_{\tau}}{\sqrt{2}c_\gamma}\times(\Delta^{\rho}_R \mu^c \tau
 + \Delta^{\rho}_L
 \tau^c\mu ) \left(H^0_2+i H_{A_1}\right) +\mathrm{h.c.}.
  \label{phigsdeay1}\ee
 This Lagrangian has the same form as that of
  the  MSSM in the limit of the CP-odd neutral Higgs having heavy mass.
 The lepton flavor conserving (LFC) part of the Lagrangian at tree level can be deduced from \cite{Dong3}.
  Using the notation in \cite{giang}, this part has the form
 \be \mathcal{L}^{FC}_{H^0\mu\tau}=
 -\left(Y_{\mu}\mu^c\mu+Y_{\tau}\tau^c\tau\right) \rho^{\prime0}+
 \mathrm{h.c.}. \label{LFlagrang}\ee
 We note that light Higgs $H^0_1$ has very suppressed LFV
 effective couplings in this case. At
 the tree level, charged leptons only couple
 to Higgs $\rho'$ and $ \sqrt{2}\rho^{\prime0}=
    \left( H^0_1
  c_{\gamma}+ H^0_2s_{\gamma}\right)+ i \left(c_{\gamma}
  H_{A_4}+s_{\gamma}
   H_{A_1}\right)$.  The LFV branching ratio of neutral Higgses $H^0$
   can be calculated through
   the  branching ratios  $\mathrm{BR}(H^0\rightarrow \tau^+\tau^-)$,
   namely,
\bea \mathrm{BR}(H^0\rightarrow
\tau^+\mu^-)&=&\mathrm{BR}(H^0\rightarrow
\tau^-\mu^+)\crn&=&\frac{1}{c_\gamma^2s_\gamma^2}\times
\left(\left|\Delta^{\rho}_L\right|^2
 + \left|\Delta^{\rho}_R\right|^2\right)
\mathrm{BR}(H^0\rightarrow \tau^+\tau^-)\crn &=&
\frac{(t^2_{\gamma}+1)^2}{t^2_{\gamma}}
\left(\left|\Delta^{\rho}_L\right|^2
 + \left|\Delta^{\rho}_R\right|^2\right)
\mathrm{BR}(H^0\rightarrow \tau^+\tau^-), \label{phigsdeay2}\eea
where $H^0= H_{A_1}, H^0_2$.

In the case of $t_{\gamma} \gg 1$, as obtaining  the
Lagrangian (\ref{FCNC2}), we  obtain a result that is the same as
that indicated in the  MSSM  for heavy neutral Higgses. We have
\be \mathrm{BR}(H^0\rightarrow \tau^+\mu^-)= t_\gamma^2
\left(\left|\Delta^{\rho}_L\right|^2
 + \left|\Delta^{\rho}_R\right|^2\right)
\mathrm{BR}(H^0\rightarrow \tau^+\tau^-).\label{phigsdeay2}\ee
 The  neutral Higgs-fermion-fermion couplings in our work are different from
 \cite{Dong3}. They are listed in
  Table \ref{table1}. We just
 consider for $H^0_1,~H^0_2$ and $H_{A_1}$.
  \begin{table}[h]
\caption{ Coupling of neutral Higgs bosons to
fermions.}\label{table1}
  \begin{tabular}{|c |c|c|c|c|}
    \hline
   Particles & Up-fermion& Down-fermion & Exotic up-quark &
   Exotic down-quark\\
    \hline
    SM  Higgs & 1 & 1 & 0 &0\\
    $H^0_1$ & $1 $ & $1 $
     &$ \mathcal{O}\left(\frac{m_W}{m_X}\right)$&$ \mathcal{O}\left(\frac{m_W}{m_X}\right) $ \\
    $H^0_2$ & $-\frac{1}{t_{\gamma}} $& $t_{\gamma}  $& $ \mathcal{O}\left(\frac{m_W}{m_X}\right)$&
    $ \mathcal{O}\left(\frac{m_W}{m_X}\right)$\\
     $H_{A_1}$ & $\frac{i}{t_{\gamma}} $& $ i t_{\gamma}  $& $ \mathcal{O}\left(\frac{m_W}{m_X}\right)$&
    $ \mathcal{O}\left(\frac{m_W}{m_X}\right)$\\
    \hline
     \end{tabular}
\end{table}
 Following this table, the couplings of light neutral
 Higgs to fermions are the same as those in the SM.
  While the CP-even  and CP-odd neutral Higgses are
 different,  they strongly couple with the down fermion with large $t_\gamma$.
  Furthermore, these two Higgses do weakly couple with exotic
 quarks of the model. They carry properties of neutral Higgses in
  the MSSM and the $\nu$MSSM shown in \cite{Dia1}. As mentioned in \cite{Dia1} and as
 detailed for example in \cite{Djou}, the coupling of these Higgses to
 $W^+W^-$ and $Z^0Z^0$ are very suppressed if  their masses are
 very heavy. For the SUSYE331, a similar case also occur for the vertex type
  of $H^0VV$ where $V$
  denotes any gauge bosons $Z,~Z',~W^{\pm},~Y^{\pm}$ or $X^0$.
  The couplings are deduced from the following term:
  \be \sum_{H^0} \left( D^{\mu}H^0\right)^\dagger \left( D_{\mu}H^0\right) \rightarrow ig^2_{V}
  V_{\mu}V^{\mu\dagger}\sum_{H^0}\left(\langle H^0\rangle H^{0\dagger}\right)
  +\mathrm{H.c.}, \label{coupliHVV}\ee
 where  $g_V$ is defined from covariant derivative
 $D_{\mu}=\partial_{\mu}+i\sum_{V} g_{V}
 V_{\mu}$. As shown in the Appendix \ref{ehiggsmaas}, the leading
 contributions to Higgses $H^0_1,~H^0_2$ and $H_{A_1}$  come only from
 the two Higgses $\rho^0$ and $\rho^{\prime0}$, the couplings with
 all gauge bosons are proportional to $g_V^2 m_W/g$.  The next
 leading contributions are  related with $\chi^0$ and $\chi^{\prime 0}$ by a factor of
 $\sqrt{\epsilon}=m_W/m_X$. Because these two Higgses contain real
 components having VEVs $w,w'\sim \frac{2m_X}{g}$, the value of the coupling
 of $H^0_1VV$ is proportional to $ g_V^2m_W/g^2$. In
 contrast, the coupling of  $H^0_2VV$ is still suppressed because
 of a factor $s_{2\gamma}<\frac{1}{t_{\gamma}}$. So in the case of our work
 the leading and next leading contributions to
 couplings $HVV$ of $\left\{H^0_{1},~H^0_2,~H_{A_1}\right\}$  to gauge
 bosons are $g_V^2 m_{W}/g\times\left\{
 \sin\gamma,0,0\right\}$ and $g_V^2
 m_{W}/g^2\times\left\{\mathcal{O}(1),s_{2\gamma},0\right\}$, respectively.
 It means that coupling of $H^0_{2}VV$  is very suppressed and
 $H_{A_1}$ does not couple to gauge boson pairs.  This is the same case as in
  the MSSM and the $\nu$MSSM. Therefore,  $H^0_2$ and $H_{A_1}$
  decay mainly to down fermions
 such as $b\bar{b}$ and $\tau\bar{\tau}$ \cite{ern}. This will lead to large LFV branching
 ratios of neutral heavy
 Higgses which can be detected by
 the LHC. A Detailed investigation  can
 be found in \cite{Dia1} for example.

\section{Conclusion}
In this work we have  concentrated  on the Higgs sector of the
SUSYE331 model. Unlike the previous  work \cite{susyec,Dong3}, by
adding two $B/\mu$-type
 terms in the soft term of the SUSYE331 model we
 have shown that
 these terms  not only guarantee the  vacuum  stability   but
 also cancel all of the tachyon Higgses appearing in the previous
 version. Especially, from the conditions of the minimum of the scalar
 potential we indicated that soft parameters and the $B/\mu$-terms in
 this model naturally favor  the order of $\mathrm{SU(3)}_L$. This is the
 property of the SUSYE331 model which does not occur in
 supersymmetric versions of the $\mathrm{SU(2)}_L\times\mathrm{
 U(1)}_Y$. Because of this,  all of three CP-odd neutral Higgses will
 get masses at least around 1 TeV. They are denoted 
 $m^2_{A_1},~m^2_{A_2}$ and $m^2_{A_3}=m^2_{A_2}+m^2_X$.
 These four Higgs states are found exactly according to the original
Higgs basis. For the neutral Higgs sector, there are four massive
Higgses in which there is one light Higgs with squared mass $
m^2_{H^0_1}\simeq m^2_Zc_{2\gamma}^2$, the same as 
in the  MSSM. Among the three other CP-even neutral heavy Higgses,
there is one exact value $m^2_{H^0_4}=m^2_{A_3}$. In the charged
Higgs sector, there is also one exact value of the charged Higgs mass, 
$m^2_{H^{\pm}_4}=m^2_{A_1}+m^2_W$. This formula suggests the
similarity of the Higgses $H^{\pm}_4$ and $H_{A_1}$ to those in the
MSSM. In summary, in the limit of large values of  the soft parameters,
$B/\mu$-type terms, $t_\gamma$ and $t_\beta$ the Higgs spectrum of
the SUSYE331 contains all Higgses carrying many properties of the
MSSM Higgs spectrum. The remaining ones characterize the SUSYE331 because
they almost relate with the $\mathrm{SU(3)}_L$ Higgses $\chi$ and
$\chi'$.  Among these Higgses, there maybe exists a charged Higgs
tachyon, unless the conditions
$\frac{(m_{A_1}^2+m_W^2)c_{2\gamma}}{m_X^2}<c_{2\beta}
<\frac{c_{2\gamma}m^2_W}{m_X^2+m^2_{A_2}}<0 $ are satisfied. They
 give  two important consequences: i) for $t_\gamma>1$
($c_{2\gamma}<0$) $t_\beta$ is larger than 1 too, ii) if the value of
$m^2_{A_1}$ is very close to the value of
$\left(\frac{c_{2\beta}}{c_{2\gamma}} m^2_X-m^2_W\right)$ there
will appear a light charged Higgs characteristic  for the existence of the
$\mathrm{SU(3)}_L$  itself, which supports the charged
Higgs searches at LHC and other colliders.

 It is emphasized that the above classification helps us to exploit  many known results 
for the MSSM to estimate the properties of the first class of Higgses
in the SUSYE331, although they seem  to be only true  at
   the tree level. For completeness  it is really necessary to study in detail  the
   effect from loop corrections because new particles will generate new
diagrams in higher order calculations. As an illustration, we
consider the LFV decays of neutral Higgs bosons to leptons in the
SUSYE331. The loop contributions to these decays were indicated in
\cite{giang}. This result does not depend on the appearance of
$B/\mu$-type terms. The calculation in this work shows that the LFV
decays of the three neutral Higgses $H_{A_1},~H^0_1$ and $H^0_2$ are
consistent with the conclusions for
 the  MSSM neutral Higgses shown in \cite{Dia1}. Here  the $H^0_1$
is the lighter CP-even neutral Higgs. It is normally identified
with SM like-Higgs. The two Higgses $H_{A_1}$ and $H^0_2$ are very
heavy Higgses with degenerate masses. Furthermore they decay
mainly to down fermions such as $b\bar{b}$ and $\tau\bar{\tau}$
leading to the enhancement of LFV branching ratios up to
$\mathcal{O}(10^{-4})$ for the   MSSM and
 the SUSYE331. This is really new and of significance for the heavy neutral
Higgses which can be checked by experiments.

\section*{Acknowledgments}
L.T. Hue would like to thank the referee of \cite{hue1} for
his/her suggestion about  this work.  This research is funded by
Vietnam National Foundation for Science and Technology Development
(NAFOSTED) under grant number 103.01-2011.63.

\appendix
\section{\label{ehiggsmaas}CP-even neutral Higgs squared mass matrix}
 We list precisely all of elements of the CP-even neutral Higgs squared
 mass  matrix as follows
 \bea m_{S11}^2&=& \frac{1}{2}
\frac{b_{\chi}}{t_{\beta}}+
\frac{g^2}{8}w^{\prime2}+\frac{g^2}{108}\left(18+t^2\right)u^2,\crn
m_{S12}^2&=&-\frac{g^2}{8}u^{\prime}w^\prime+\frac{g^2}{108}\left(18+t^2\right) u w,\nonumber \\
m_{S13}^2&=&-\frac{1}{2}b_{\chi} -\frac{g^2}{8} ww^{\prime} -
\frac{g^2}{108}\left(18+t^2\right)uu^\prime,\crn
m_{S14}^2&=&\frac{g^2}{8}u^{\prime}w-\frac{g^2}{108}\left(18+t^2\right)uw^\prime, \nonumber \\
m^2_{S15}&=&-\frac{g^2}{108}\left(9+2t^2\right)uv,\hspace*{2.5cm}
 m^2_{S16}=\frac{g^2}{108}\left(9+2t^2\right)uv^\prime \nonumber \\
m^2_{S22}&=&\frac{b_{\chi}}{2
t_{\beta}}+\frac{g^2}{8}u^{\prime2}+\frac{g^2}{108}\left(18+t^2\right)w^2,\crn
m^2_{S23}&=&\frac{g^2}{8}u^{\prime}w-\frac{g^2}{108}\left(18+t^2\right)u^\prime w, \nonumber \\
m^2_{S24}&=&-\frac{b_{\chi}}{2}-\frac{g^2}{8}uu^\prime-\frac{g^2}{108}\left(18+t^2\right)w
w^\prime,\crn m^2_{S25}&=&-\frac{g^2}{108}\left(9+2t^2\right)w v,
\hspace*{2.5cm} m^2_{S26}=\frac{g^2}{108}\left(9+2t^2\right) w
v^\prime,\crn
 m^2_{S33}&=&\frac{b_{\chi}}{2
t_{\beta}}+\frac{g^2}{108}\left(18+t^2\right)u^{\prime2}+\frac{g^2}{8}w^{ 2}\nonumber \\
m^2_{S34}&=&\frac{g^2}{108}\left(18+t^2\right)u^\prime
w^\prime-\frac{g^2}{8}uw,\hspace*{1.4cm}
m^2_{S35}=\frac{g^2}{108}\left(9+2t^2\right) u^\prime v, \nonumber
\\ m^2_{36}&=&-\frac{g^2}{108}\left(9+2t^2\right) u^\prime v^\prime, \crn
 m^2_{S44}&=&
\frac{b_{\chi}}{2t_{\beta}}+\frac{g^2}{108}\left(18+t^2\right)w^{\prime2}+\frac{g^2}{8}u^{2},\nonumber \\
m^2_{S45}&=&\frac{g^2}{108}\left(9+2t^2\right)
vw^\prime,\hspace*{2.5cm}
m^2_{46}=-\frac{g^2}{108}\left(9+2t^2\right)w^\prime
v^\prime ,\nonumber \\
m^2_{S55}&=&\frac{b_{\rho}}{2
t_{\gamma}}+\frac{g^2}{54}\left(9+2t^2\right)v^2,\hspace*{2.5cm}
m^2_{S56}=- \frac{g^2}{54}\left(9+2t^2\right)v v^\prime, \nonumber \\
m^2_{S66}&=&\frac{b_{\rho}}{2
t_{\gamma}}+\frac{g^2}{54}\left(9+2t^2\right) v^{\prime 2}.
\nn\eea

To estimate contributions from  original  Higgs basis $S_i$ to
physical Higgs basis we  do a rotation of the squared mass matrix
(\ref{higgstrunghoa}) with the rotation  $C$ represented as follows:
\bea C=\left(%
\begin{array}{cccccc}
 s_{\beta}  & 0 &c_{\beta}&0  &0& 0 \\
 -c_{\beta}&0  & s_{\beta} & 0 & 0 &0  \\
  0&0  & 0 & 0 & -c_{\gamma}  & s_{\gamma} \\
   0 & 0 & 0 & 0 &s_{\gamma} & c_{\gamma}\\
   0& s_{\alpha} & 0 & c_{\alpha} & 0 & 0 \\
   0& c_{\alpha} & 0 & -s_{\alpha}& 0 & 0 \\
\end{array}%
\right),\label{rotaC1}\eea
 where
\bea
t_{2\alpha}&=&t_{2\beta}\times\frac{m^2_{A_2}+m^2_{Z'}}{m^2_{A_2}-m^2_{Z'}},
\hs  \mathrm{and}\crn \;
\frac{s_{2\alpha}}{s_{2\beta}}&=&-\frac{m^2_{A_2}+m^2_{Z'}}
{\sqrt{c^2_{2\beta}\left(m^2_{A_2}-m^2_{Z'}\right)^2+s^2_{2\beta}\left(m^2_{A_2}+m^2_{Z'}\right)^2}}
\simeq -\frac{m^2_{H_3}+m^2_{H_4}}{-m^2_{H_3}+m^2_{H_4}}.\crn
\label{dnalpha}\eea
 Because $s_{2\beta}>0$ we have $s_{2\alpha}<0$. The sign of
 $c_{2\alpha}$ depends on the quantity $m^2_{A_2}-m^2_{Z'}$.
 Because of this we have $\pi/2<\alpha<3\pi/2$.

  After this rotation, we keep only large
contributions to the
 squared mass matrix which are  proportional  to $m_X^2$, $m_W m_X$
  and $m^2_W$ in the non-diagonal elements of the matrix. Then
 we have
{\small \bea  M^2_{H^0}&=& C.M^2_{6S}.C^T\crn
= m^2_X&\times& \left(%
\begin{array}{cccccc}
  0 & 0 & 0 & 0 & 0 & 0 \\
  0, &k_2+1, & 0&
  0 & 0 & 0 \\
   0 & 0, & k_1+\frac{s^2_{2\gamma}\epsilon}{4c^2_W-1}, & \frac{2\epsilon\sin4\gamma}{4c^2_W-1},&
     -\frac{2\sqrt{\ep}s_{2\gamma}\cos(\beta+\alpha)}{4c^2_W-1},
     &\frac{2\sqrt{\ep}s_{2\gamma}\sin(\beta+\alpha)}{4c^2_W-1} \\
  0 & 0 & \frac{2\epsilon\sin4\gamma}{4c^2_W-1}, & \frac{4c^2_{2\gamma}\epsilon}{4c^2_W-1}, &
 - \frac{2\sqrt{\ep}c_{2\gamma}\cos(\beta+\alpha)}{4c^2_W-1}, &
  \frac{2\sqrt{\ep}c_{2\gamma}\sin(\beta+\alpha)}{4c^2_W-1} \\
     0 & 0 & -\frac{2\sqrt{\ep}c_{2\gamma}\cos(\beta+\alpha)}{4c^2_W-1},
    &-\frac{2\sqrt{\ep}s_{2\gamma}\cos(\beta+\alpha)}{4c^2_W-1}, & M^2_{55}, &0 \\
     0 & 0 &  \frac{2\sqrt{\ep}c_{2\gamma}\sin(\beta+\alpha)}{4c^2_W-1},
    & \frac{2\sqrt{\ep}s_{2\gamma}\sin(\beta+\alpha)}{4c^2_W-1}, & 0, &M^2_{66} \\
\end{array}%
\right) \crn \label{massnH2}\eea}
 where
 \bea M^2_{55}\times m_X^2&=&m^2_{Z'}
\cos^2(\beta+\alpha)+ m^2_{A_2}\sin^2(\beta-\alpha),\crn
 M^2_{66}\times
 m_X^2&=&m^2_{Z'}\sin^2(\beta+\alpha)+m^2_{A_2}\cos^2(\beta-\alpha).\label{massnH255}
 \eea
In this new basis, all  non-diagonal elements of the squared mass
matrix are of the order $\mathcal{O}(\sqrt{\ep})$ or $\mathcal{O}(\ep)$.
So we can use this basis of the Higgses to represent mass eigenstates of
heavy Higgses. In particular,  these states are related with the
originals by
 \bea
 m^2_{H^0_5} &=& m^2_{A_2}+m^2_X;\hspace*{6.0cm}
 H^0_5= -c_{\beta} S_1+s_{\beta} S_3, \crn
 m^2_{H^0_2}&= & m^2_{A_1}+ \mathcal{O}(m^2_W);
\hspace*{5.5cm}   H^0_2= -c_{\ga} S_5 + s_{\ga} S_6, \crn
 m^2_{H^0_3} &=&m^2_{Z'}
\cos^2(\beta+\alpha)+ m^2_{A_2}\sin^2(\beta-\alpha)+
\mathcal{O}(m^2_W)
 ;\;
H^0_3= s_{\alpha} S_2+c_{\alpha} S_4
 \crn m^2_{H^0_4}&=&m^2_{Z'}\sin^2(\beta+\alpha)+m^2_{A_2}\cos^2(\beta-\alpha)+
\mathcal{O}(m^2_W);
 \;
   H^0_4= c_{\alpha} S_2-s_{\alpha} S_4.\crn \label{heavynHstate}\eea
  In addition, we have a massless state $ H^\prime =  s_{\beta} S_1+
  c_{\beta}S_3$ eaten by $X^0$ boson. For the light Higgs
   we can see from matrix (\ref{massnH2}) the diagonal element
 $\left( M^2_{H^0}\right)_{44}
 =\frac{4 m^2_Wc^2_{2\gamma}}{4c^2_W-1}=\frac{4c^2_W}{4c^2_W-1}m^2_Z c^2_{2\gamma}$ is different from the
  eigenvalue of $m^2_Zc^2_{2\gamma}$ predicted in
  (\ref{nlHiggs1}). This is because of the non-diagonal elements
  in the matrix (\ref{massnH2}) which are proportional to
  $\sqrt{\ep}$. They can cause corrections order of $\ep\times
  m^2_X\simeq m^2_W$ to all Higgs masses and affect directly the mass of the light
  Higgs. For example, we consider the case of large $t_\gamma$
  and $t_\beta$. This means that $\gamma, \beta\rightarrow \pi/2$
  and $\sin4\gamma=s_{2\gamma}\rightarrow 0,~c_{2\gamma}\rightarrow-1$.
   Furthermore,  because  $\alpha$ is defined in (\ref{dnalpha}) and  $m^2_{A_2}<m^2_X$ as
  chosen  in a numerical investigation we obtain  $\pi/2<\alpha<\pi$ and $\alpha\rightarrow
  \pi/2$. Inserting these values
   into (\ref{massnH255}) we have
  $m^2_{55}\rightarrow m^2_{Z'}\cos^2(\beta+\alpha)$ and  the largest contributions to Higgs masses
   from the non-zero diagonal elements are only
   $\left( M^2_{H^0}\right)_{45}=\left(
  M^2_{H^0}\right)_{54}=2
  \sqrt{\ep}c_{2\gamma}\cos(\beta+\alpha)/(4c^2_W-1)$.  We then take a
  rotation with a tiny angle $\eta$ defined by
  $$\tan2\eta=-\frac{4\sqrt{\epsilon}\cos2
  \gamma\cos(\beta+\alpha)}{(4c^2_W-1)M^2_{55}-4c^2_{2\gamma}\epsilon}.$$
 The light Higgs mass  now is
 $$ m^2_{H^0_1}\simeq m^2_X\times\left[\frac{c^2_{2\gamma}\epsilon}{4c^2_W-1}- \frac{1}{m^2_{55}}
  \times\left(\frac{2\sqrt{\ep}c_{2\gamma}\cos(\beta+\alpha)}{4c^2_W-1}\right)^2\right]\simeq m^2_{Z}$$
 as predicted. In this case the mass eigenvalue of the light Higgs
  has the form $ H^0_1= s_{\ga} S_5 + c_{\ga} S_6+
  \mathcal{O}(\frac{m_W}{m_X})\times\left( s_{\alpha} S_2+c_{\alpha}
  S_4\right)$.

 In general, the dominant contributions to mass eigenstate of
 light Higgs is $ H^0_1= s_{\ga} S_5 + c_{\ga} S_6$.  The next
 contributions to this eigenstate and other heavy Higgses
 $H^0_2,~H^0_3$ and $H^0_4$ are all proportional to a factor of
 $m_W/m_X\simeq0.03$. This contribution to $H^0_2$ is more
 suppressed because of a factor $s_{2\gamma}\sim
 \frac{1}{t_{\gamma}}$. So these contributions can be ignored in many
  investigations such as LFV decays of neutral Higgses.

\section{\label{chiggsmass}Charged Higgs squared mass matrix}
The non-zero elements of charged Higgs squared mass matrix
are listed as follows:
 \bea m^2_{\chi^-\chi^{+}}&=& \frac{4b_{\chi}}{g^2}+
\left(w^{2\prime}+u^{2\prime}\right) +(v^2-v^{\prime2}),\crn\hs
m^2_{\chi^-\chi^{\prime+}}&=&-
\frac{4b_{\chi}}{g^2}-t_{\beta}\left(w^{2\prime}+u^{2\prime}\right),\crn
m^2_{\chi^{\prime-}\chi^{\prime+}}&=&
\frac{4b_{\chi}}{g^2}+(u^2+w^2)-(v^2-v^{2\prime}),\hs
 m^2_{\rho^-_1\rho_1^+}=\frac{4b_{\rho}}{g^2t_{\gamma}}+(u^2-u^{\prime2})+v^{\prime2},\crn
m^2_{\rho^-_1\rho_2^+}&=& uw-u^\prime w^\prime,\hs
m^2_{\rho_2^-\rho_2^+}= \frac{4b_{\rho}}{g^2t_{\gamma}}+
v^{\prime2}+(w^2-w^{\prime2}),\hs\crn
 m^2_{\rho_1^{\prime
-}\rho_1^{\prime+}}&=&\frac{4b_{\rho}t_{\gamma}}{g^2}+v^2+(u^{\prime
2}-u^2),\hs m^2_{\rho^{\prime-}_1
\rho^{\prime+}_2}=-m^2_{\rho^{-}_1\rho^{+}_2}, \nonumber \\
m^2_{\rho_2^{-\prime}\rho_2^{+\prime}}&=&\frac{4b_{\rho}t_{\gamma}}{g^2}+v^2+(w^{2\prime}-w^2).
\eea
In the limit of $u~,u'\rightarrow 0$, the matrix has a simpler
form, and after taking a rotation this matrix by a transformation
 $\mathcal{C}_{H^{\pm}}$ with
 {\footnotesize\bea  \mathcal{C}_{H^{\pm}} =\left(%
\begin{array}{cccccc}
  -\frac{s_{\beta} m_X }{\sqrt{m_X^2+m_W^2}} & -\frac{c_{\beta} m_X }{\sqrt{m_X^2+m_W^2}} &
  0& \frac{s_{\gamma} m_W}{\sqrt{m_X^2+m_W^2}}  & 0 & \frac{c_{\gamma} m_W}{\sqrt{m_X^2+m_W^2}} \\
  0 & 0 & s_{\gamma} & 0 & c_{\gamma} & 0 \\
  0 & 0 & c_{\gamma} & 0 & -s_{\gamma} & 0 \\
  0 & 0 & 0 & c_{\gamma} & 0 & -s_{\gamma} \\
  c_{\beta} & 0 & -s_{\beta} & 0 &0 & 0 \\
  \frac{s_{\beta} m_X }{\sqrt{m_X^2+m_W^2}} & \frac{c_{\beta} m_X }{\sqrt{m_X^2+m_W^2}} &
  0& \frac{s_{\gamma} m_W}{\sqrt{m_X^2+m_W^2}}  & 0 & \frac{c_{\gamma} m_W}{\sqrt{m_X^2+m_W^2}} \\
\end{array}%
\right),\label{RcHiggs1} \eea } we get
{\footnotesize \bea
&&\mathcal{C}_{H^{\pm}}M^2_{6\mathrm{charged}}\mathcal{C}_{H^{\pm}}^{\dagger}=\crn
&&\left(%
\begin{array}{cccccc}
  0 & 0 & 0 & 0 & 0 & 0 \\
  0 & 0 & 0 & 0 & 0 & 0 \\
  0 & 0 & m^2_{A_1}+m^2_W & 0 & 0 & 0\\
  0 & 0 & 0 & m^2_{A_1}+m^2_W-c_{2\beta}c_{2\gamma}m^2_X, & s_{2\gamma}s_{2\beta} m_W m_X,  &
  -s_{2\gamma} c_{2\beta} m_X m_U  \\
 0 & 0 & 0 & s_{2\gamma}s_{2\beta} m_W m_X, &m^2_{A_2}-c_{2\beta}c_{2\gamma} m^2_W+m^2_X
  &-c_{2\gamma} s_{2\beta} m_W m_U\\
 0 & 0 & 0 & -s_{2\gamma} c_{2\beta} m_X m_U & -c_{2\gamma} s_{2\beta} m_W m_U & c_{2\beta}c_{2\gamma} (m_W^2 + m_X^2) \\
\end{array}%
\right).\crn \label{chargeh2}\eea}
From this  and $ \left(
H^+_6,~H^+_5,~H^+_4,~H^{\prime+}_3,~H^{\prime+}_2,~H^{\prime+}_1\right)^T=
\mathcal{C}_{H^{\pm}} (\chi^+,~\chi^{\prime+},~\rho^{+}_1,
~\rho^{+}_2,~\rho^{\prime+}_1, ~\rho^{\prime+}_2)^T$ we easily see
that there are two Goldstone bosons $H^{\pm}_5$ which  are eaten
by $W^{\pm}$ and two  massive Higgses, $H^{\pm}_4$, with  masses
arising  mainly from $\rho$ and $\rho'$.
\section{\label{correctionNH}Corrections to lightest neutral Higgs
mass}
To illustrate the contribution from the loop corrections
to lightest neutral Higgs mass, we use the same  simplest
estimation as done in the  MSSM \cite{ellis} namely
\begin{itemize}
\item We choose $\beta\rightarrow \frac{\pi}{2}$,
$\gamma\rightarrow \frac{\pi}{2}$ and  $u\rightarrow 0$. This
limit leads to $w'\rightarrow 0$,~ $w\rightarrow W=2 m_X/g$,
 $v'\rightarrow
0$ and $v \rightarrow V= 2m_W/g$. This choice is consistent with
$b_{\rho} \rightarrow 0 $,~ $b_{\chi} \rightarrow 0 $,
$\frac{1}{4}\mu_{\rho}+m^2_{\rho'}\rightarrow \infty$, and
$\frac{1}{4}\mu_{\chi}+m^2_{\chi'}\rightarrow \infty$. Hence the
antitriplets $\chi^\prime, \rho^\prime$ can be integrated out
when we consider the symmetry breaking  of $SU(3)_L \times
U(1)_X$. For convenience we define the new parameters such as
 \bea m^2_1 &=&
\frac{1}{4}\mu^2_{\chi}+m^2_{\chi}, \hs m^2_2=\frac{1}{4}\mu^2_{\rho}+m^2_{\rho}, \\
 \frac{S_5+v}{\sqrt{2}}&\rightarrow& \phi_2, \hs \mathrm{and}\hs  \frac{S_2+W}{\sqrt{2}}\rightarrow \phi_1.\eea
 With these conventions, the superpotential at the tree level can be written as
\bea  V_{\mathrm{SUSYE331}}\rightarrow V_0=
m_2^2\phi^2_2+m_1^2\phi^2_1+ \frac{9g^2+2g^{\prime2}}{54} \left[k
\phi^4_1 -\phi_1^2\phi_2^2+ \phi^4_2 \right],\label{Vlimit1}\eea
 where $t^2\equiv(g'/g)^2=18s_{W}^2/(3-4s^2_W)$ and $k=(18+t^2)/[2(9+2t^2)]=c^2_W$.  The
 tree level minimization gives
 \bea  \left. \frac{\partial V_0}{\partial
 \phi_1}\right|_{\phi_1=W/\sqrt{2},~\phi_2=V/\sqrt{2}}&=& 0 \rightarrow
 m^2_1=-\frac{9+2t^2}{27}\left( 2 k m_X^2-m^2_W\right),\crn
  \left. \frac{\partial V_0}{\partial
 \phi_2}\right|_{\phi_1=W/\sqrt{2},~\phi_2=V/\sqrt{2}}&=& 0 \rightarrow
 m^2_2=\frac{9+2t^2}{27}\left( m_X^2-2 m^2_W\right).
 \label{minimuml1}
 \eea
 The mass Lagrangian at tree level related with term
 $\left(\frac{\partial^2 V_0}{\partial \phi_i \partial
 \phi_i} \right)$ can be written as
 \be  \mathcal{L}_{\mathrm{mass}}=-\frac{4(9+2t^2)m^2_X}{27} \left(%
\begin{array}{cc}
  \phi_1, & \phi_2 \\
\end{array}%
\right) \left(%
\begin{array}{cc}
  2k & -\epsilon' \\
  -\epsilon' & 2\epsilon'^2 \\
\end{array}%
\right) \left(%
\begin{array}{c}
  \phi_1 \\
  \phi_2 \\
\end{array}%
\right)\label{l0mass}\ee
 with $\epsilon' = \sqrt{\epsilon}=\frac{m_W}{m_X} \ll 1$ which we can use as a perturbative
 parameter. The lightest mass eigenvalue is
 \be m^2_{0h}= \frac{2(9+2t^2)m^2_X}{27}\left( k+\epsilon'^2-\sqrt{( k-\epsilon'^2)^2+\epsilon'^2}\right). \ee
Using the approximation
$\sqrt{(k-\epsilon'^2)^2+\epsilon'^2}\simeq
 k-\epsilon'^2+\frac{\epsilon'^2}{2k}$ we obtain  $ m^2_{0h}\simeq
m^2/c^2_W \simeq m^2_Z$, being consistent with the result shown in
appendix \ref{ehiggsmaas}.This result confirms that the
VEV of $\chi$ gives a tiny contribution  to the lightest neutral
 Higgs mass. Now we construct the effective potential for neutral Higgs at the one-loop level.
  We concentrate on terms related with only $\phi_2$ which give the largest contribution to the mass of
  the lightest CP-even neutral Higgs.Let us remind the reader the role of the triplets $\chi$
    and $\rho$ in generating mass for quarks. The Yukawa interactions containing $\chi, \rho$ are given
    by
    \bea \mathcal{L}^{Y}_{u}=-\frac{1}{3}
    \left[\kappa_{4 \al i}Q_{\al L}d^c_{iL}\chi+\kappa'_{4 \al \beta}Q_{\alpha L}d^{\prime
    c}_{\beta L}\chi+\kappa_{3\alpha i}Q_{\alpha L}u^{c}_{iL}\rho+\kappa'_{3\alpha i}Q_{\alpha L}u^{\prime
    c}_{iL}\rho
    \right]. \label{masupq1}\eea We choose
    $ \kappa_{4 \al i} \rightarrow 3 \kappa_4\delta_{\al i}$,  $ \kappa_{3\alpha i} \rightarrow -
    3 y_{3\alpha}\delta_{\alpha i}$  ($ y_c\equiv y_{32}$, $ y_t\equiv y_{33}$) and ignore the mixing of top and exotic u-quarks.
     Therefore the mass of top quark is $m_t=y_t v/\sqrt{2}$.

The masses of sfermions in the SUSYE331 were analyzed  in
    \cite{sfSUSYE331}. In this work with the assumption of the Yukawa term the
    largest supersymmetric  contributions to the masses of the two left and right stops are the same
    and equal to $y_t \phi_2$. For the simplest case, we also assume that the
    contribution from the soft term for each left or right stop
    is $m^2_{\tilde{q}}$. All contributions for stop quark coming from the D-term are
    ignored. This assumption is similar to that given in
 \cite{ellis}. Hence, the  squared masses of the top quark and stop have the form
\be m^2_t=y^2_t \phi_2^2, \hs m^2_{\tilde{t}}=y^2_t \phi_2^2
+m^2_{\tilde{q}}.\label{topmass1}\ee
\end{itemize}
 The full one-loop potential now is
 \be V(Q)=V_0(Q)+\Delta V_1(Q), \label{fspotential1}\ee
 where
 \be \Delta V_1(Q)=\frac{1}{64\pi^2}\mathrm{Str}\left[ \mathcal{M}^4\left(
 \ln\frac{\mathcal{M}^2}{Q^2}-c\right)\right].
 \label{dfspotential1}\ee
 Here $\mathcal{M}^2$ is the field-dependent generalized squared
 mass matrix and the supertrace is defined as
 \be
 \mathrm{Str}f(\mathcal{M}^2)=\sum_{i}(-1)^{2J_i}(2J_i+1)f(m^2_i)\label{supertrace1}\ee
 with $J_i$ is the spin of the field having mass $m_i$. We
take contribution only from top quarks and stops, namely
 \be  \Delta V_1(Q)=\frac{3}{16\pi^2}\left[2 m^4_{t} \ln\left(\frac{m^2_{\tilde{t}}}{m^2_{t}}\right)
 +\left(m^4_{\tilde{t}}-m^4_{t}\right)\left(
 \ln\frac{m^4_{\tilde{t}}}{Q^2}-c\right)\right].\label{dfspotential1}\ee
   From the Eq. (\ref{topmass1}) we have
   \be
   \frac{\partial(m^2_{t})}{\partial(\phi_1)}=\frac{\partial(m^2_{\tilde{t}})}{\partial(\phi_1)}=0, \ee
   \be \frac{\partial(m^2_{t})}{\partial(\phi_2)}=\frac{\partial(m^2_{\tilde{t}})}{\partial(\phi_2)}=2y_t^2
   \phi_1. \ee
 This leads to the consequence that  $\frac{\partial \Delta
   V_1}{\partial\phi_1}=0$. The minimal condition is
   equivalent to the following equation:
   \bea  && \left.\frac{\partial\left( V_0+ \Delta
   V_1\right)}{\partial\phi_2}\right|_{\phi_1=W/\sqrt{2},~\phi_2=V/\sqrt{2}}=0\crn
   &&\rightarrow \left(m^2_{\tilde{t}}
   -m^2_{t}\right)\left(\ln\frac{m^2_{\tilde{t}}}{\hat{Q}}-c\right)=-\left(m^2_{\tilde{t}}
   -m^2_{t}\right)-2 m^2_{t} \ln\frac{m^2_{\tilde{t}}}{m^2_{t}}.
   \label{minicon1}\eea
  Because of the Eq. (\ref{minicon1}) one can obtain
\bea
\frac{\partial^2V(\phi_1,\phi_2)}{\partial\phi^2_2}=\frac{\partial^2V_0}{\partial\phi^2_2}+
\frac{3g^2}{8\pi^2}\frac{m^4_t}{m^2_W}\ln\left(\frac{m^4_{\tilde{t}}}{m^4_t}\right).
\label{huo} \eea The last term in Eq. (\ref{huo}) is the
correction from the one-loop effective potential. The squared
mass matrix for neutral Higgs is obtained as follows:
 \be  \frac{1}{2}\left[ \frac{\partial^2V(\phi_1,\phi_2)}{\partial\phi_i ~\partial
 \phi_j}\right]_{\phi_1=W/\sqrt{2},~\phi_2=V/\sqrt{2}}.
 \label{massM}\ee
Diagonalizing the matrix found from Eq. (\ref{massM}) we find 
the formula of the lightest squared mass  as follows:
\be m^2_{h}= \frac{2}{3-4s^2_W}\left(c^2_W m^2_X+\Delta
+m^2_W-\sqrt{( c^2_W m^2_X-\Delta-m_W^2)^2+m^2_Xm^2_W}\right)
\label{lightm3}\ee
 with $\Delta=
 \frac{3g^2(3-4s^2_W)}{64\pi^2}\frac{m^4_t}{m^2_W}\ln\left(\frac{m^4_{\tilde{t}}}{m^4_t}\right)$. In case of
 $\Delta\sim \mathcal{O}(m^2_W)\ll m^2_{X}$ we obtain  $m^2_{h}\simeq m^2_Z+ \frac{3g^2}{16\pi^2}
 \frac{m^4_t}{m^2_W}\ln\left(\frac{m^4_{\tilde{t}}}{m^4_t}\right)$.
 This result is the same  as
that in the  MSSM.


\begin{thebibliography}{999}
\bibitem{higgsdicovery1}The ATLAS Collaboration, Phys.Lett. B716 (2012) 1,
arXiv:1207.7214.
 \bibitem{higgsdicovery2} The CMS Collaboration, G. Aad et al, Phys. Lett.
B 716 (2012) 30, arXiv:1207.7235.
\bibitem{lowmH}S. Heinemeyer, O. St{\aa}l, G. Weiglein, Phys. Lett. B710 (2012)
 201; M. Carena, S. Heinemeyer, O. St{\aa}l, C.E.M. Wagner, G.
Weiglein, arXiv: hep-ph/1302.7033.
\bibitem{NonSMHiggs1}L. Hall, D. Pinner and J. Ruderman, JHEP 1204 (2012) 131, arXiv:hep-ph/1112.2703;
\bibitem{NonSMHiggs2}A.  Arbey, M. Battaglia, A. Djouadi, F. Mahmoudi and J. Quevillon, Phys. Lett. B 708
(2012) 162, arXiv: hep-ph/1112.3028; P. Draper, P. Meade, M. Reece
and D. Shih, Phys. Rev. D 85 (2012) 095007, arXiv:
hep-ph/1112.3068.
\bibitem{CERNreport} The LHC Higgs Cross Section Working Group Collaboration (S. Heinemeyer (eds.) et
al.), arXiv: hep-ph/1307.1347.
\bibitem{guido}G.  Altarelli, arXiv: hep-ph/1308.0545.
\bibitem{Djouadi2}A. Djouadi, Phys. Rept. 457 (2008) 1.
hep-ph/0503172.
\bibitem{Djou} A. Djouadi, Phys. Rept. 459 (2008) 1.
\bibitem{Martin}S. P. Martin, hep-ph/9709356.
\bibitem{degrassi} G. Degrassi, S. Heinemeyer, W. Hollik, P. Slavich and G. Weiglein, Eur. Phys. J. C 28
(2003) 133 [arXiv:hep-ph/0212020].
\bibitem{331m} F. Pisano and V. Pleitez, Phys. Rev.  D {\bf 46}, 410 (1992);
P. H. Frampton, Phys. Rev. Lett. {\bf 69}, 2889 (1992); R. Foot,
O. F. Hernandez, F. Pisano and V. Pleitez, Phys. Rev. D {\bf 47},
4158 (1993).
\bibitem{331r} M. Singer, J. W. F. Valle and J. Schechter, Phys.
Rev. D {\bf 22}, 738 (1980);    R. Foot, H. N. Long, T. A. Tran,
Phys. Rev. D 50 (1994) 34R,
 [arXiv: hep-ph/9402243];
J. C. Montero, et al., Phys. Rev. D 47 (1993) 2918; H. N. Long,
Phys. Rev. D 54 (1996) 4691;  H. N. Long, Phys. Rev. D 53 (1996)
437;
 H. N. Long,   Mod. Phys. Lett. {\bf A13}, (1998) 1865.
\bibitem{ecq}F. Pisano, Mod. Phys. Lett A {\bf 11}, 2639 (1996);
A. Doff and F. Pisano, Mod. Phys. Lett. A {\bf 14}, 1133 (1999);
C.  A.  de S. Pires and O. P. Ravinez, Phys. Rev. D {\bf 58},
035008 (1998); C. A. de S. Pires, Phys. Rev. D {\bf 60}, 075013
(1999); P. V. Dong and H. N. Long, Int. J. Mod. Phys. A {\bf 21},
6677 (2006).
\bibitem{longvan} H. N. Long and V. T. Van, J. Phys. G {\bf 25}, 2319
(1999).
 \bibitem{haihiggs} P. V. Dong, H. N. Long, D. T. Nhung, D. V.   Soa, Phys. Rev. D {\bf 73} (2006) 035004;
 P. V. Dong and H. N. Long, Adv. High Energy Phys.
{\bf 2008}, 739492 (2008), [arXiv:0804.3239(hep-ph)].
\bibitem{Ferreira}J. G. Ferreira, Jr, P. R. D. Pinheiro, C. A. de S. Pires,
P. S. Rodrigues da Silva, Phys. Rev. D {\bf 84} (2011) 095019.
\bibitem{higgseconom} P.  V.  Dong, H. N. Long, D. V. Soa, Phys. Rev. D {\bf 73}  (2006) 075005.
\bibitem{susyec} P. V.  Dong, D. T. Huong, M. C. Rodriguez, H. N. Long, Nucl. Phys. B 772 (2007) 150.
\bibitem{s331r} J. C. Montero, V. Pleitez, M. C. Rodriguez, Phys. Rev. D 70 (2004)
075004; S.  Sen, Phys. Rev. D76 (2007) 115020.
\bibitem{susyrm331}D. T. Huong, L. T. Hue, M. C. Rodriguez, H. N.
Long, Nucl. Phys. B 870 (2013) 293-322; J. G. Ferreira, C. A. de
S. Pires, P.  S. Rodrigues da Silva, A. Sampieri, arXiv:
hep-ph/1308.0575.
\bibitem{Dong3}P. V. Dong, D. T. Huong, N. T. Thuy and H. N.
Long, Nucl. Phys. \textbf{B 795} (2008) 361; arXiv:0707.3712 --
\bibitem{inflasusye331}D. T. Huong, H. N. Long, J. Phys. G38 (2011)
015202, arXiv: hep-ph/1004.1246; Do T. Huong, H. N. Long, Phys.
Atom. Nucl. 73 (2010) 791, arXiv: hep-ph/0807.2346. ; H. N. Long,
Adv. Stud.Theor. Phys. 4 (2010) 173, arXiv: hep-ph/0710.5833.
\bibitem{susymaspectrum} D. T. Huong, H. N. Long, JHEP 0807 (2008) 049, arXiv:
hep-ph/0804.3875.

\bibitem{sfSUSYE331} P. V. Dong, Tr. T.
Huong, N. T. Thuy, H. N. Long, JHEP 0711 (2007) 073, arXiv:
hep-ph/0708.3155.
\bibitem{giang}P. T. Giang, L. T. Hue, D. T.
Huong and H. N. Long, Nucl. Phys. B 864 (2012) 85.
\bibitem{hue1} L.  T. Hue, D. T. Huong, H. N. Long, Nucl. Phys. B 873 (2013)
207, arXiv: hep-ph/1301.4652.
\bibitem{coutinho} Y. A. Coutinho, V. S. Guimar\~{a}es and A.  A.
Nepomuceno, arXiv/Hep-ph: 1304.7907.
\bibitem{assa} K.  A.  Assamagan,  A. Deandrea, P. Delsart, Phys.Rev. D67 (2003)
035001.
\bibitem{lfvlhc}S.  Davidson,  P.  Verdier, Phys. Rev. D {\bf 86} (2012)
11170, arXiv: hep-ph/1211.1248;

\bibitem{lfvMSSM}S. Kanemura, K. Matsuda, T. Ota, T. Shindou, E. Takasugi and K.Tsumura
Phys. Lett. \textbf{B}599(2004)83, hep-ph/0406316; S. Kanemura, T.
Ota, T. Shindou and K.Tsumura, Phys. Rev. D {\bf 73} (2006)
016006, hep-ph/0505191; K. S. Babu and C. Kolda, Phys. Rev. Lett.
\textbf{89} (2002) 241802, arXiv: hep-ph/ 0206310;
 M.  Arana-Catania, E.
Arganda, M. J. Herrero, arXiv: hep-ph/1304.3371;
\bibitem{Anna1} A.  Brignoble,  A.  Rossi, Phys. Lett. \textbf{B 566} (2003)
217, arXiv:hep-ph/0304081;  A. Brignole, A. Rossi, Nucl. Phys.  B
701 (2004),3-53; arXiv: hep-ph/ 0404211.
\bibitem{Dia1} J. L.  Diaz-Cruz, D. K.  Ghosh and S. Moretti, Phys. Lett. B 679 (2009)
376.
\bibitem{abb} Abbiendi G et al.
[ALEPH and DELPHI and L3 and OPAL and The LEP working group for
Higgs boson searches Collaborations], \emph{Search for Charged
Higgs bosons: Combined Results Using LEP Data}, arXiv/hep-ex:
1301.6065.
\bibitem{Barat1}R.  Barate et al. [LEP Working Group for Higgs boson searches and
ALEPH and DELPHI and L3 and OPAL Collaborations], Phys. Lett. B
565 (2003) 61 [arXiv:hep-ex/0306033].
\bibitem{ern}E. Arganda, J. L. Diaz-Cruz, A. Szynkman, Eur. Phys. J. C 73 (2013)
2384. \bibitem{ellis} J. Ellis, G. Ridolfi and F. Zwirner,
Phys. Lett. B 257(1991) 83.
\bibitem{corr} H. E. Haber and  R. Hempfling, Phys. Rev.
Lett. 66 (1991) 1815; A. Brignole, Phys. Lett. B 281 (1992) 284.

\end{thebibliography}
\end{document}